\documentclass[apj]{emulateapj}
\usepackage{amsmath}
\usepackage{graphicx}
\usepackage{natbib}
\usepackage{longtable}
\usepackage{url}

\newcommand{\hbeta}{H{$\beta$}}

\newcommand{\civ}{C{\sevenrm IV}}
\newcommand{\mgii}{Mg{\sevenrm II}}
 
 \font\sevenrm=cmr7 scaled 1000

\begin{document}

\title{Quasar Black Hole Mass Estimates in the Era of Time Domain Astronomy}
\shorttitle{Variability-based Mass Estimates}
\shortauthors{Kelly et al.}

\author{Brandon C. Kelly\altaffilmark{1}, Tommaso Treu\altaffilmark{1}, Matthew Malkan\altaffilmark{2}, Anna Pancoast\altaffilmark{1}, Jong-Hak Woo\altaffilmark{3}}

\altaffiltext{1}{Department of Physics, Broida Hall, University of California, Santa Barbara, CA 93106-9530}
\altaffiltext{2}{Department of Astronomy, 430 Portola Plaza, Box 951547, University of California, Los Angeles, CA 90095-1547}
\altaffiltext{3}{Department of Physics and Astronomy, Seoul National University, Seoul, 151-742, Republic of Korea}

\begin{abstract}

We investigate the dependence of the normalization of the high-frequency part of the X-ray and optical power spectral densities (PSD) on black hole mass for a sample of 39 active galactic nuclei (AGN) with black hole masses estimated from reverberation mapping or dynamical modeling. We obtained new Swift observations of PG 1426+015, which has the largest estimated black hole mass of the AGN in our sample. We develop a novel statistical method to estimate the PSD from a lightcurve of photon counts with arbitrary sampling, eliminating the need to bin a lightcurve to achieve Gaussian statistics, and we use this technique to estimate the X-ray variability parameters for the faint AGN in our sample. We find that the normalization of the high-frequency X-ray PSD is inversely proportional to black hole mass. We discuss how to use this scaling relationship to obtain black hole mass estimates from the short time-scale X-ray variability amplitude with precision $\sim 0.38$ dex. The amplitude of optical variability on time scales of days is also anti-correlated with black hole mass, but with larger scatter. Instead, the optical variability amplitude exhibits the strongest anti-correlation with luminosity. We conclude with a discussion of the implications of our results for estimating black hole mass from the amplitude of AGN variability.

\end{abstract}


\section{Introduction}
\label{s-intro}

The ability to measure black hole mass, $M_{BH}$, is of fundamental importance for active galactic nuclei (AGN) research. When combined with luminosity, the black hole mass gives an estimate of the Eddington ratio, providing two of the most important quantities for studies of accretion physics and the growth of supermassive black holes.  Estimates of $M_{BH}$ and $L / L_{Edd}$ are important for understanding the physics of accretion flows because many aspects of accretion physics at fixed radius scale with $M_{BH}$ and accretion rate \citep[e.g.,][]{Frank2002a}, and the Eddington ratio is related to the accretion rate. Estimates of $M_{BH}$ and $L / L_{Edd}$ are important for understanding the growth of supermassive black holes because the mass distribution directly probes the buildup of the black hole population, while the Eddington ratio distribution is related to the instantaneous growth rate of the black hole population. Moreover, the energy output from accretion onto the black hole scales with the black hole's mass, and is thought to be an important ingredient in models of galaxy formation and evolution \citep[e.g.,][]{Di-Matteo2005a,Bower2006a,Croton2006b,Hopkins2006a}.

It is well accepted that currently the most reliable method for estimating $M_{BH}$ is through modeling the spatially resolved stellar and gas dynamics in the host galaxy's nucleus. However, for most AGN this is not feasible, both due to the presence of the active nucleus and due to the strong requirements on the  angular resolution of the nucleus. Maser kinematics offers another avenue for estimating $M_{BH}$ \citep[e.g.,][]{Greene2010a}, but are rare. For the vast majority of AGN reverberation mapping \citep[e.g.,][]{Blandford1982a,Peterson1993a} provides the most reliable means of estimating $M_{BH}$. In general, masses derived from reverberation mapping are only good to within a multiplicative scaling factor that varies for each AGN and depends on the geometry of the broad line region, the importance of radiation pressure, and other potential sources for systematic error \citep[e.g.,][]{Krolik2001a}. Typically an average scaling factor is used, which is estimated by assuming the reverberation mapping estimates of $M_{BH}$ lie on the $M_{BH}$--$\sigma_*$ relationship \citep[e.g.,][]{Onken2004a,Woo2010a,Graham2011a,Park2012a,Woo2013a,Grier2013b}, although efforts are underway to derive this geometric factor from the data \citep{Pancoast2011a,Brewer2011a,Pancoast2012a} or from direct dynamical modeling of the gas in the nuclear regions \citep{Hicks2008a}. Using this average scaling factor, $M_{BH}$ estimates from reverberation mapping are generally considered to be good to within a factor of a few, as estimated from the scatter in the masses about the $M_{BH}$--$\sigma_*$ relationship \citep[e.g.,][]{Collin2006a,Woo2010a}.

Unfortunately, the spectroscopic monitoring campaigns required of reverberation mapping are expensive to carry out, and thus far mass estimates derived using this technique have only been obtained for $\sim 50$ low redshift AGN \citep[e.g.,][]{Peterson2004a,Bentz2009b,Denney2010a,Barth2011b,Barth2011a,Grier2012a}. Instead, scaling relationships are used to obtain mass estimates for large numbers of AGN over a broad range in redshift and luminosity. Such scaling relationships are based on correlations between $M_{BH}$ and some feature, or set of features, that are easier to measure. The scaling relationships are calibrated to masses derived from either stellar and gas dynamical modeling or from reverberation mapping. The scatter in the masses derived from the scaling relationships at fixed $M_{BH}$ defines the statistical error in the mass estimates with respect to the masses derived from dynamical modeling or reverberation mapping.

The most common scaling relationships used to estimate $M_{BH}$ are those based on the host galaxy properties \citep[e.g.,][]{McLure2002a,Merritt2001a,Tremaine2002a,Haring2004a,Graham2007a,McConnell2013a}, with the luminosity or stellar velocity dispersion of the bulge being the most common proxies for $M_{BH}$, or those based on the AGN luminosity and broad emission line widths \citep[e.g.,][]{Wandel1999a,McLure2002b,Vestergaard2006a,Shen2011a}; scaling relationships involving the AGN radio and X-ray luminosities \citep[e.g.,][]{Merloni2003a,Falcke2004a} or X-ray variability \citep[e.g.,][]{Czerny2001a,Nikolajuk2004a,Gierlinski2008a,Kelly2011a} are also occasionally used. By far the most common scaling relationship used to estimate $M_{BH}$ in AGN is the one involving the broad emission lines. Broad line mass estimates have been used in studies of AGN accretion physics \citep[e.g.,][]{Shemmer2006a,Kelly2008a,Trump2011a,Davis2011a}, black hole growth \citep[e.g.,][]{McLure2004a,Netzer2007a,Vestergaard2009a,Kelly2010a,Trakhtenbrot2011a,Shen2012a,Kelly2013a}, and evolution in scaling relationships between the host galaxy and $M_{BH}$ \citep[e.g.,][]{Treu2004a,Peng2006a,Treu2007a,Woo2008a,Bennert2010a,Merloni2010a}.

The various scaling relationships for estimating $M_{BH}$ have their advantages and disadvantages. Moreover, there is considerable potential for systematic error in these scaling relationships, and employing mass estimates derived from more than one scaling relationship is important for assessing the impact of unknown systematics on the scientific conclusions. The broad line mass estimates can only be obtained for Type 1 AGN, and thus are not applicable to AGN that are either obscured or diluted by their host galaxy. Moreover, several studies have indicated that the statistical scatter in these mass estimates for AGN in a narrow luminosity range is smaller than the $\sim 0.4$ dex scatter seen about the reverberation mapping masses \citep[e.g.,][]{Kollmeier2006a,Fine2008a,Shen2008a,Steinhardt2010a,Shen2012a,Kelly2013a}, which may be indicative of a bias in these mass estimates as a function of luminosity \citep{Shen2010a,Shen2012a,Shen2013a}, at least when the line width is estimated using the FWHM. And finally, the broad line mass estimates are usually calculated based on the \hbeta, \mgii, or \civ\ emission lines. The mass estimates based on \hbeta\ are considered the most secure, primarily due to the fact that most of the reverberation mapping analysis is based on \hbeta, but there is still considerable uncertainty regarding the systematics of the \mgii\ and \civ\ lines. There is almost no reverberation mapping data involving the \mgii\ line \citep[e.g.,][]{Woo2008b}. The \civ\ line is often considered the most problematic, as it is thought to arise from the base of a wind launched from the accretion disk, and thus may have a significant non-virial component \citep[e.g.,][]{Baskin2005a,Richards2011a}. Indeed, while the FWHM for \hbeta\ and \mgii\ correlate, the FWHM between \mgii\ and \civ\ exhibit only a weak correlation \citep[e.g.,][]{Shen2012b} likely due to the presence of a non-variable component in the emission line \citep{Denney2012a}. However, the mass estimates do correlate for \hbeta\ and \civ\ when one uses the second moment of the \civ\ line obtained from high $S/N$ spectra \citep{Denney2013a}. And finally, the AGN studied via reverberation mapping occupy a narrower range of parameter space with respect to their emission line properties than does the general population \citep{Richards2011a}, and it is unclear if systematic errors become significant when extrapolating beyond this region of parameter space.

In contrast to the broad line mass estimates, mass estimates derived from the host galaxy properties can be obtained for AGN that are obscured or diluted by their host. However, the host galaxy properties are difficult to impossible to measure for bright AGN, and the broad line mass estimates must be used in this case. Moreover, much recent work has shown that the host galaxy scaling relationships are more complicated than originally thought, and depend on the host galaxy morphology \citep[e.g.,][]{Hu2008a,Gultekin2009a,Greene2010a,Kormendy2011a,Graham2012a,McConnell2013a}. In addition, there is considerable uncertainty regarding the form of these scaling relationships beyond the local universe. While many studies have found evidence that the normalization in the scaling relationships evolves \citep[e.g.,][but see \citep{Lauer2007a,Shen2010a,Schulze2011a} for cautionary notes]{Peng2006a,Treu2007a,Merloni2010a,Bennert2011a,Canalizo2012a,Schramm2013a}, there is still considerable uncertainty in whether the slope or scatter also evolve. Indeed, many models connecting the growth of supermassive black holes and their host galaxies predict evolution in the scaling relationships \citep[e.g.,][]{Robertson2006a,Croton2006a,Volonteri2009a}.

X-ray variability provides a third scaling method for estimating $M_{BH}$ that is competitive with the broad emission line and host galaxy techniques, and can overcome some of their respective disadvantages. In fact, recent work has suggested that scaling relationships involving X-ray variability provide mass estimates with $\sim 0.3$ dex precision \citep[e.g.,][]{Zhou2010a,Kelly2011a,Ponti2012a}, which is comparable to mass estimates derived from the $M_{BH}$--$\sigma_*$ relationship for ellipticals \citep{Gultekin2009a} and better than mass estimates derived from the broad emission lines \citep[$\sim 0.4$ dex,][]{Vestergaard2006a}. Even before reliable $M_{BH}$ estimates were available it was suggested that X-ray variability properties should scale with $M_{BH}$ \citep[e.g.,][]{Barr1986a,Wandel1986a,McHardy1988a,Green1993a}. Recent work has also found trends between the optical variability properties of AGN and $M_{BH}$ \citep[e.g.,][]{Collier2001a,Kelly2009a,MacLeod2010a}, although they are not as well established as the X-ray trends. However, the investigation of such trends will provide an important foundation for interpreting and utilizing the variability information from current and planned time-domain optical surveys, such as those obtained by Pan-STARRS and the Large Synoptic Survey Telescope (LSST).

The origin of the X-ray variability scaling relationship is a dependence of the X-ray power spectral density (PSD) on black hole mass. The X-ray PSDs of both galactic black holes and supermassive black holes are similar within the uncertainty caused by the poorer data quality of AGN lightcurves \citep[e.g.,][]{McHardy2004a}, especially for galactic black holes in the `high-soft' state. AGN X-ray PSDs are typically well-modeled by a bending power law form, $P(f) \propto 1 / f^{\alpha}$, where the PSD slope is $\alpha \sim 1$ down to some high-frequency break $f_H$, and then steepens to $\alpha \sim 2$ at frequencies $f > f_H$. Many authors have found a significant anti-correlation between $f_H$ and $M_{BH}$ \citep[e.g.,][]{Uttley2002a,Markowitz2003a,McHardy2004a,Uttley2005b,Kelly2011a,Gonzalez-Martin2012a}, which extends all the way down to galactic black holes. The time scale corresponding to the break frequency scales with $M_{BH}$ and luminosity approximately as $\tau_H \propto M_{BH}^2 / L$ \citep{McHardy2006a,Kording2007a}, which implies $\tau_H \propto M_{BH} / \dot{m}$ where $\dot{m}$ is the accretion rate relative to Eddington. In principle the X-ray variability time scale can be used to estimated $M_{BH}$, but in practice it is hard to measure, requiring a lightcurve that is well sampled on time scales near $\tau_H$. 

A related scaling relationship between the amplitude of the PSD at frequencies $f > f_H$ and $M_{BH}$ shows considerable promise as a mass estimator \citep{Hayashida1998a,Czerny2001a,Nikolajuk2004a,Gierlinski2008a,Kelly2011a,McHardy2013a}. The normalization of the high-frequency PSD can be measured by fitting a power-law to the periodogram (i.e., the empirical power spectrum) \citep[e.g.,][]{Green1993a,Czerny2001a,Gierlinski2008a} or by fitting a stochastic process to the measured lightcurve \citep{Kelly2009a,Kelly2011a}. Alternatively, it is common to use the excess variance \citep[e.g.,][]{Nandra1997a} as a probe of the high-frequency PSD. The excess variance is the fractional variance in a X-ray lightcurve after subtracting off the variance due to the Poisson noise from counting photons. It is related to the amplitude of the high-frequency PSD because it is the integral of the PSD over the time scales probed by a lightcurve. Several groups have found a significant anti-correlation between the excess variance and $M_{BH}$ \citep{Lu2001a,Bian2003a,Nikolajuk2004a,Papadakis2004a,ONeill2005a,Nikolajuk2006a,Miniutti2009a,Zhou2010a,Caballero-Garcia2012a,Ponti2012a,McHardy2013a}. Results obtained from fitting the normalization of the high-frequency PSD and the excess variance imply that these scaling relationships can be used to derive mass estimates with $\sim 0.2$--$0.3$ dex precision relative to the mass estimates obtained from reverberation mapping or through dynamical modeling \citep{Zhou2010a,Kelly2011a}. Because the mass estimates derived from reverberation mapping themselves are uncertain, it is possible that the true scatter in the high-frequency X-ray variability relative to $M_{BH}$ may be even smaller. Moreover, motivated by upcoming massive time-domain optical surveys, such as those performed by LSST, it is worth investigating if similar trends with $M_{BH}$ hold for the optical variability as well; the existence of such trends could potentially enable black hole mass estimates for millions of quasars. 

\begin{figure}
	\includegraphics[scale=0.5,angle=00]{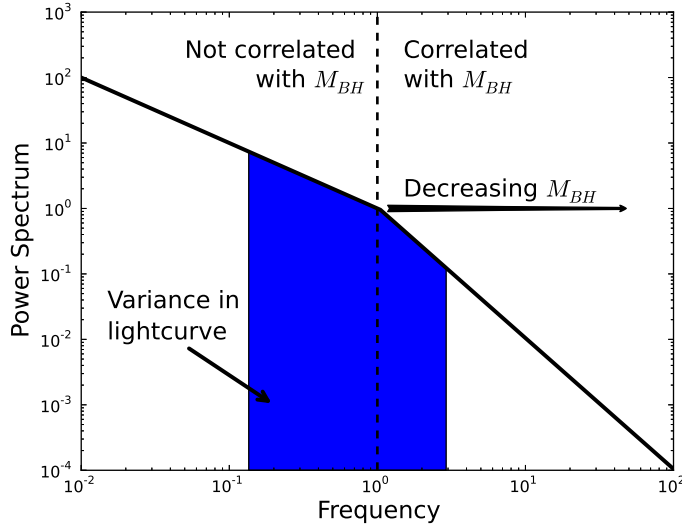}
	\caption{Illustration of how the variance in a lightcurve relates to the PSD features, neglecting the window function caused by sampling a continuous time process; note that the units are arbitrary. Both the break frequency and the PSD amplitude above the break depend on $M_{BH}$. The variance of a lightcurve is given by the integral of the PSD over the frequencies corresponding to the time scales probed by the lightcurve. When the lightcurve is probing time scales corresponding to frequencies above the break, then the variance is correlated with $M_{BH}$. However, when the lightcurve is probing time scales below the break, the correlation with $M_{BH}$ weakens. Because the break frequency is in general unknown, this can cause problems when using the lightcurve variance as a $M_{BH}$ estimator. In this work we describe a statistical technique to always measure the PSD amplitude above the break frequency, which can then be used to estimate $M_{BH}$.
		\label{f-illust_psd}}
\end{figure}

While there is significant potential in high-frequency X-ray variability as a mass estimator, in practice it can be difficult to measure. The periodogram is distorted by the finite, and possibly irregular, sampling of the lightcurve. This can cause biases in the derived high-frequency normalization when fitting the empirical PSD, and Monte Carlo techniques are required to correct for this \citep{Uttley2002a}. The excess variance, while easy to measure, is also problematic. In order for the excess variance to be proportional to the normalization of the high-frequency PSD, the length of the lightcurve must not be significantly longer than the break time scale, $\tau_H$, or must be broken up into smaller segments. If the break frequency is not known, there is no guarantee that this requirement is satisfied. The fact that the break frequency increases with decreasing $M_{BH}$ further complicates this issue. Moreover, because the expected value of the excess variance is the integral of the PSD over the time scales probes by the lightcurve, it will depend on the length and sampling of the lightcurve, and thus must be normalized to some fiducial lightcurve length; failure to correct for the effects of sampling can lead to biases in the inferred excess variance \citep[e.g.,][]{Pessah2007a,Allevato2013a}. These issues regarding sampling the PSD above the break frequency are illustrated in Figure \ref{f-illust_psd}. And finally, the excess variance is a noisy estimate, exhibiting large changes when measured in non-overlapping intervals from a stochastic lightcurve \citep{Vaughan2003a}. Monte Carlo methods need to be used to assess the bias and uncertainty in the excess variance estimates. These issues can inhibit the widespread use of the X-ray variability as a black hole mass estimator.

The techniques developed by \citet[][hereafter KBS09 and KSS11, respectively]{Kelly2009a,Kelly2011a} that derive the high-frequency PSD by fitting a stochastic process to the lightcurve do not suffer from these distortions caused by the sampling pattern. Moreover, KBS09 and KSS11 derive the likelihood function for the PSD parameters, including the break frequency $f_H$ and the normalization of the PSD at $f > f_H$, under the assumption that the high frequency PSD falls off as $P(f) \propto 1 / f^2$. Thus in these models the amplitude of high-frequency variability is a parameter which can be estimated via maximum-likelihood or Bayesian methods, providing an optimal use of the information in the data and rigorous uncertainties on the parameters. This is an advantage over the excess variance or empirical PSD techniques, as it provides more precision in the mass estimates than does the excess variance, and it does not require expensive Monte Carlo simulations.

Essentially all methods that have been used to estimate the high-frequency PSD have assumed that the measurement errors on the background-subtracted lightcurve are normally distributed. However, this assumption becomes inaccurate for faint X-ray sources, in which case it may be necessary to use large time bins to ensure approximate Gaussian statistics, thus reducing the amount of variability information. In addition, this assumption becomes inaccurate when the background count rate becomes comparable to the source count rate. In order to enable X-ray variability-based mass estimates for large numbers of AGN it is necessary to develop statistical techniques that can be applied directly to the X-ray counts.

Motivated by these issues, we have set out in this paper to first develop a method for estimating the normalization of the high-frequency PSD that works directly on a time series of X-ray counts without employing Gaussian approximations, thus providing tools to obtain X-ray variability based $M_{BH}$ estimates in practice. After doing this, we seek to derive scaling relationships between the amplitude of high-frequency X-ray variability and $M_{BH}$ using 34 AGN with $M_{BH}$ estimated from reverberation mapping or dynamical modeling; this is the largest sample yet of its kind used for investigating this scaling relationship. These scaling relationships not only provide the calibration needed to use the amplitude of high-frequency X-ray variability as a mass estimator, but also provide valuable empirical constraints on the physics of the accretion flows. And finally, using this same set of AGN we investigate whether the optical variability also shows significant trends with $M_{BH}$ that might enable it to be used as a mass estimator, as well as provide empirical constraints on the physics of the accretion flows. 

\section{Data}
\label{s-data}

Our sample consists of those AGN that have black hole masses estimated from reverberation mapping or dynamical modeling. All values of $M_{BH}$ derived from reverberation mapping were recalculated from the virial factors assuming a value of $f = 4.38$ \citep{Grier2013b}, where $f$ is the proportionality constant between $M_{BH}$ and the virial factor. Using this data set as a starting point, we searched the literature for archival X-ray and optical lightcurves. In addition, we obtained a new X-ray lightcurve for PG 1426+015 from Swift. Our sample is summarized in Table \ref{t-sample}. The $M_{BH}$ estimates were taken from the references given in the Table, and the bolometeric luminosities were taken from \citet{Vasudevan2009a} or \citet{Woo2002a}.

\begin{deluxetable*}{lcccccc}
\tabletypesize{\scriptsize}
\tablewidth{0pt}
\tablecaption{Sample Properties \label{t-sample}}
\tablehead{
Name
& $z$
& $\log M_{BH}$
& $\log L\tablenotemark{a}$
& Reference\tablenotemark{b}
& X-ray Data
& Optical Data? \\
&
& $M_{\odot}$ 
& ergs s$^{-1}$
&
&
&
}
\startdata
3C 120 & 0.03301 & 7.71 $\pm$ 0.04 & 45.3 & 1 & XMM & Y \\
3C 390.3 & 0.0561 & 9.04 $\pm$ 0.05 & 45.2 & 1 & XMM & Y \\
Ark 120 & 0.032713 & 7.98 $\pm$ 0.09 & 45.3 & 1 & XMM & Y \\
Fairall 9 & 0.047016 & 8.29 $\pm$ 0.09 & 44.8 & 2 & KSS11\tablenotemark{c} & Y \\
Mrk 766 & 0.012929 & 6.10 $\pm$ 0.29 & 44.4 & 1 & KSS11 & Y \\
Mrk 110 & 0.03529 & 7.32 $\pm$ 0.14 & 45.1 & 1 & XMM & Y \\
Mrk 279 & 0.030451 & 7.47 $\pm$ 0.06 & 45.0 & 1 & XMM & Y \\
Mrk 290 & 0.029577 & 7.27 $\pm$ 0.07 & 44.4 & 4 & \ldots & Y \\
Mrk 335 & 0.025785 & 7.03 $\pm$ 0.11 & 45.1 & 2 & XMM & Y \\
Mrk 50 & 0.023433 & 7.41 $\pm$ 0.07 & 43.8 & 9 & XMM & N \\
Mrk 509 & 0.034397 & 7.95 $\pm$ 0.02 & 45.2 & 1 & XMM & N \\
Mrk 590 & 0.026385 & 7.47 $\pm$ 0.09 & 43.8 & 1 & XMM & Y \\
Mrk 6 & 0.018813 & 8.02 $\pm$ 0.04 & 44.4 & 10 & XMM & N \\
Mrk 79 & 0.022189 & 7.89 $\pm$ 0.14 & 44.3 & 1 & XMM & Y \\
Mrk 817 & 0.031455 & 7.77 $\pm$ 0.07 & 45.0 & 1 & XMM & Y \\
NGC 3227 & 0.003859 & 7.18 $\pm$ 0.17 & 42.5 & 3 & KSS11 & Y \\
NGC 3516 & 0.008836 & 7.47 $\pm$ 0.04 & 43.5 & 1 & XMM & Y \\
NGC 3783 & 0.00973 & 7.26 $\pm$ 0.08 & 44.1 & 1 & KSS11 & Y \\
NGC 4051 & 0.002336 & 6.32 $\pm$ 0.09 & 42.6 & 1 & KSS11 & Y \\
NGC 4151 & 0.003319 & 7.65 $\pm$ 0.05 & 42.8 & 5 & XMM & Y \\
NGC 4395 & 0.001064 & 5.44 $\pm$ 0.13 & 40.7 & 6 & Chandra & N \\
NGC 4593 & 0.009 & 6.94 $\pm$ 0.08 & 43.7 & 1 & XMM & Y \\
NGC 5548 & 0.017175 & 7.76 $\pm$ 0.06 & 44.3 & 1 & XMM & Y \\
NGC 6814 & 0.005214 & 7.19 $\pm$ 0.06 & 43.9 & 1 & XMM & Y \\
NGC 7469 & 0.016317 & 7.30 $\pm$ 0.13 & 44.8 & 1 & XMM & Y \\
PG 0026+129 & 0.142 & 8.48 $\pm$ 0.11 & 45.4 & 2 & \ldots & Y \\
PG 0052+251 & 0.15445 & 8.45 $\pm$ 0.09 & 45.8 & 2 & XMM & Y \\
PG 0804+761 & 0.1 & 8.72 $\pm$ 0.05 & 45.9 & 2 & XMM & Y \\
PG 0953+414 & 0.2341 & 8.32 $\pm$ 0.09 & 46.5 & 2 & XMM & Y \\
PG 1226+023 & 0.158839 & 8.94 $\pm$ 0.09 & 47.1 & 2 & XMM & Y \\
PG 1229+204 & 0.06301 & 7.80 $\pm$ 0.07 & 44.9 & 1 & XMM & Y \\
PG 1307+085 & 0.155 & 8.52 $\pm$ 0.12 & 45.6 & 2 & XMM & Y \\
PG 1411+442 & 0.0896 & 8.01 $\pm$ 0.14 & 45.4 & 1 & XMM & Y \\
PG 1426+015 & 0.08657 & 9.16 $\pm$ 0.08 & 45.6 & 1 & Swift & Y \\
PG 1613+658 & 0.129 & 8.33 $\pm$ 0.2 & 45.9 & 2 & \ldots & Y \\
PG 1617+175 & 0.112438 & 8.65 $\pm$ 0.2 & 45.5 & 1 & \ldots & Y \\
PG 1700+518 & 0.292 & 8.77 $\pm$ 0.10 & 46.6 & 2 & \ldots & Y \\
PG 2130+099 & 0.062977 & 7.93 $\pm$ 0.06 & 45.0 & 7 & XMM & Y \\
Zw 229-015 & 0.027879 & 6.9 $\pm$ 0.09 & 43.8 & 8 & Suzaku & N \\
\enddata

\tablenotetext{a}{The bolometeric luminosity.}
\tablenotetext{b}{Reference for the $M_{BH}$ value.}
\tablenotetext{c}{The values from \citet{Kelly2011a} were used for these sources.}
\tablerefs{
(1) \citet{Grier2013b}
(2) \citet{Peterson2004a}
(3) \citet{Davies2006a}
(4) \citet{Denney2010a}
(5) \citet{Onken2007a}
(6) \citet{Peterson2005a}
(7) \citet{Grier2008a}
(8) \citet{Barth2011a}
(9) \citet{Barth2011b}
(10) \citet{Grier2012a}
}
\end{deluxetable*}

\subsection{X-ray Data}
\label{s-x-ray}

\subsubsection{Archival Data}
\label{s-archival}

Our X-ray sample consists of those AGN with X-ray variability parameters derived from KSS11 or archival X-ray data from XMM, Chandra, or Suzaku. In addition, we obtained new Swift observations of PG 1426+015. This source did not have sufficient archival X-ray lightcurves for analysis, but is important because it has the highest mass black hole out of AGN with mass estimates obtained from reverberation mapping or dynamical modeling. We did not include in our analysis IC 4329A and PG 1211+143, which \citet{Peterson2004a} deem to have unreliable $M_{BH}$ estimates derived from their reverberation mapping analysis. For each AGN not already studied by KSS11, we only use the data from a single X-ray observatory. We do this because we derive the fractional variability information from the counts for a single source, and we want to avoid any contribution to the derived variability caused by differences in the instrument calibrations, effective areas, etc. We primarily use the XMM data because it has the largest effective area, and thus should have the highest signal-to-noise for variability studies. The two exceptions were NGC 4395 and Zw 229-15. For NGC 4395 we use the Chandra data to avoid confusion with nearby X-ray sources, and for Zw 229-15 we use a 167 ksec Suzaku observation because it is much longer than the 29 ksec XMM observation. For Fairall 9, NGC 3227, NGC 3783, NGC 4051, and Mrk 766 we use the Sup-OU process parameters derived by KSS11 from flux-calibrated RXTE and XMM lightcurves; the RXTE lightcurves were constructed by \citet{Sobolewska2009a}.

The 2--10 keV counts were extracted from the XMM/PN, Chandra/ASIS, and Suzaku/XIS observations using the standard reduction routines from the XMM SAS, CIAO, and the HEASoft suite, respectively. Background counts were extracted from a nearby region. Because our analysis method requires a good estimate of the background rate (see \S~\ref{s-counts}), the background region was chosen to be much larger than the source region when possible, typically $\approx 5$--$10$ times larger. The XMM lightcurves were binned every 48 s, the lightcurve for NGC 4395 from Chandra was binned every 16.2 s, and the lightcurve for Zw 229-15 from Suzaku was binned every 160 s. For Zw 229-15 we created our final lightcurve by combining the counts from the two front-illuminated detectors, XIS-0 and XIS-3.

The sources PG 1613+658 and PG 1617+175 both have archival X-ray data but were not used in our analysis. The XMM observation for PG 1613+658 was dominated by strong background, while PG 1617+175 is X-ray weak and serendipitously observed by Chandra $\approx 12'$ off-axis.  Consequently, the PG 1617+175 observation was too faint to perform a meaningful variability analysis.

\subsubsection{New Swift Observations of PG 1426+015}
\label{s-swift}

\begin{figure}
	\includegraphics[scale=0.4,angle=90]{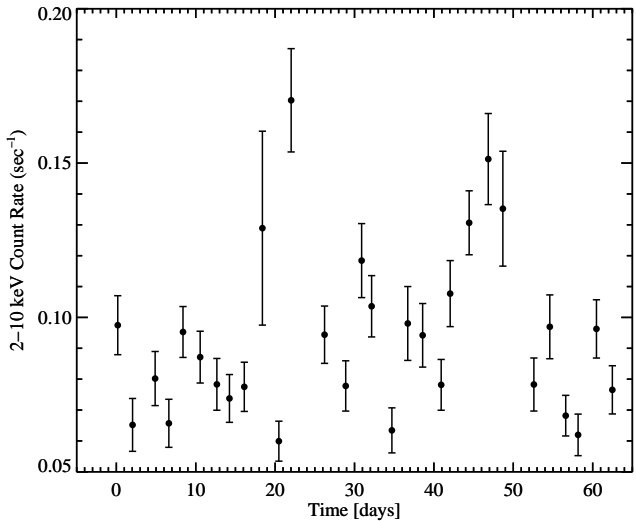}
	\caption{\emph{Swift}-XRT 2--10 keV lightcurve for PG 1426+015.
	\label{f-pg1426_lightcurve}}
\end{figure}

We obtained 30 \emph{Swift} \citep{Gehrels2004a} observations of PG 1426+015 every couple of days between 2013 April 2 and 2013 June 3. The typical exposure per visit was $\sim 2$ ksec. The OBSID associated with our monitoring campaign is 91700 with segments 01--30. The \emph{Swift}-XRT \citep{Burrows2005a} event files were created using the \emph{Swift} analysis tool \texttt{xrtpipeline} version 0.12.6. We constructed our 2--10 keV lightcurve by extracting the photon-counting mode events within a circular region centered on the source with a radius of $50"$ using \texttt{xselect} version 2.4. The background counts were extracted from a large annular region centered on the source with outer radius $377"$ and inner radius $142"$. The extracted lightcurves were corrected using the \emph{Swift} analysis tool \texttt{xrtlccorr}, which corrects the lightcurves for telescope vignetting, the point spread function, and bad pixels or columns falling within the extraction region. We compared our derived lightcurve with that generated by the UK Swift Science Data Centre XRT light curve repository\footnote{\url{www.swift.ac.uk/xrt\_lcurves/}} \citep{Evans2007a,Evans2009a} and found that they were very similar. The \emph{Swift}-XRT lightcurve for PG 1426+015 is shown in Figure \ref{f-pg1426_lightcurve}.

\subsection{Optical Data}
\label{s-optical}

Most of the optical continuum lightcurves used in this study are presented in \citet{Peterson2004a} and were obtained from the International AGN Watch\footnote{Data obtained as part of the International AGN Watch projects are available at \url{http://www.astronomy.ohio-state.edu/~agnwatch}} projects, the Lovers of Active Galaxies (LAG) campaign, the Ohio State monitoring program \citep{Peterson1998a}, and the Wise Observatory/Steward Observatory monitoring program \citep{Kaspi2000a}. These data are further discussed in \citet{Peterson2004a}. For Mrk 290, Mrk 817, NGC 3227, and NGC 3516 we used the lightcurves presented in \citet{Denney2010a}, and for Mrk 6, PG 2130+099, 3C120, and Mrk 335 we used the lightcurves presented in \citet{Grier2012a}. The optical lightcurves for Mrk 766 and NGC 6814 were taken from \citet{Walsh2009a}, who obtained these lightcurves as part of the Lick AGN Monitoring Project \citep[LAMP,][]{Bentz2009b,Barth2011b}.

\section{New Methods for Quantifying Variability from a Lightcurve of Counts}
\label{s-statmodel}

In this section we describe our extension of the stochastic modeling of KBS09 and KSS11 to lightcurves of counts. We first briefly review the Ornstein-Uhlenbeck (OU) process, and the superposition of Ornstein-Uhlenbeck processes (sup-OU), providing the necessary mathematical background. Further details can be found in KBS09 and KSS11. 

KBS09 and KSS11 derived the likelihood function of these processes under the assumption of Gaussian errors, providing a means of performing statistical inference on the parameters; \citet{Kozowski2010a} provide an alternative algorithm for calculating the likelihood function for the OU process. In this section we derive the likelihood function when the data consists of a time series of counts, and describe how to perform statistical inference on the OU and sup-OU process parameters from count data. Because we take a Bayesian approach in this work, we base our inference on an MCMC sampler which generates both random draws of the OU or sup-OU process parameters and of the count rates from their joint probability distribution, conditional on the observed counts.

\subsection{The Ornstein-Uhlenbeck Process}
\label{s-ou}

KBS09 introduced the OU process as a statistical model for the optical flux variations of AGN, and showed that it provides a good description of AGN optical variability on time scales of $\sim $ days to $\sim $ years within the data quality of ground-based observations. Subsequent work has confirmed their results \citep{Kozowski2010a,MacLeod2010a,Zu2013a,Andrae2013a}, although high-quality Kepler lightcurves show a steeper PSD slopes than expected from the OU process model on time scales $\lesssim 3$ months \citep{Mushotzky2011a}.

An OU process, $X(t)$, is described by the following stochastic differential equation:
\begin{equation}
	dX(t) = -\omega_0 (X(t) - \mu)dt + \varsigma dW(t), \ \ \ \omega_0, \varsigma > 0. \label{eq-ou}
\end{equation}
The term $dW(t)$ is a Gaussian white noise process. More general L\'{e}vy noise processes may be used in place of the Gaussian white noise process \citep[e.g.,][]{Raknerud2012a}, although statistical inference is more complicated in that case. The parameters of the OU process are the characteristic frequency, $\omega_0$, the stationary mean $\mu$, and the amplitude of driving noise, $\varsigma$. Assuming an initial condition of $X(t=0) = X_0$, the solution to Equation (\ref{eq-ou}) for $t > 0$ is
\begin{equation}
	X(t) = \mu + (X_0 - \mu) e^{-\omega_0 t} + \varsigma \int_{0}^{t} e^{-\omega_0 (t - s)} dW(s). \label{eq-ousol}
\end{equation}
It is clear from Equation (\ref{eq-ousol}) that in the absence of the driving noise (i.e., $\varsigma=0$), $X(t)$ decays back to its mean value with an $e$-folding time scale $\tau_0 = 1 / \omega_0$. The role of the driving noise is to introduce stochasticity into the lightcurve, which persistently and randomly `kicks' the lightcurve about its mean value. It is because of this interpretation that many astronomical researchers have referred to the OU process as a "damped random walk".

The stationary autocovariance function of the OU process is
\begin{equation}
	R_{OU}(t-s) \equiv {\rm cov}(X(t), X(s)) = \frac{\varsigma^2}{2\omega_0} e^{-\omega_0|t-s|} \label{eq-oucov}.
\end{equation}
Here, ${\rm cov}(x,y)$ denotes the covariance between the random variables $x$ and $y$. The stationary variance of the process is $\varsigma^2 / (2\omega_0)$, obtained by setting $t = s$ in Equation (\ref{eq-oucov}). The power spectral density (PSD) of the OU process is obtained from the Fourier transform of the autocovariance function:
\begin{eqnarray}
	P_{OU}(\omega) & = & \frac{1}{2\pi} \int_{-\infty}^{\infty} e^{-i\omega_0 t} R_{OU}(t)\ dt \nonumber \\
				 & = & \frac{\varsigma^2}{2\pi} \frac{1}{\omega_0^2 + \omega^2} \label{eq-oupsd}.
\end{eqnarray}
The PSD of an OU process is a Lorentzian centered at zero; it is flat for frequencies $\omega \ll \omega_0$ and decays as $P_{OU}(\omega) \propto \varsigma^2 / \omega^2$ for frequencies $\omega \gg \omega_0$. A flat PSD is indicative of a white noise process, implying that $\tau_0$ represents the time scale beyond which the variations in the lightcurve become uncorrelated. Hence, under the OU process model $\tau_0$ may be interpreted as the "relaxation", or "characteristic" time scale of the lightcurve, and marks the location of the knee in the PSD. The form of the autocovariance function also suggests this interpretation of $\omega_0 = 1 / \tau_0$. 

The interpretation of the term $\varsigma$, on the other hand, is not as obvious. The parameter $\varsigma$ controls the normalization of the high-frequency ($1 / \omega^2$) part of the PSD, and is thus proportional to the amplitude of short-time scale variability. Thus, $\varsigma^2$ may be interpreted as controlling the amplitude of variability on time scales $\Delta t \ll \tau$. In addition, we note that the units of $\varsigma^2$ are ${\rm rms}^2 / {\rm time}$. This, combined with Equation (\ref{eq-ou}), suggests that $\varsigma^2$ may be interpreted as the rate at which stochastic variability power is introduced into the lightcurve. Therefore, we will refer to $\varsigma^2$ as the `rate of stochastic variability power' (RSVP).

Under the assumption that $dW(t)$ is a Gaussian white noise process, the OU process itself is Gaussian. In this case, it is possible to derive the likelihood function for an OU process sampled at times $t_1 < t_2 < \ldots < t_n$. Denote the sampled values of the OU process as ${\bf x} = [x_1, \ldots, x_n]^T$, where ${\bf x}^T$ is the transpose of the vector ${\bf x}$. The likelihood function is the probability of observing ${\bf x}$ given the OU process parameters\footnote{Technically, we also condition on the time values as well. While these may also be random, we assume that their probability distribution is independent of the OU process parameters and ${\bf x}$. For simplicity we drop the dependence on $t_1,\ldots,t_n$ in our notation.}. Because the OU process is Markovian, the probability distribution of the next flux value only depends on the current flux value. Therefore, we can factor the joint probability distribution of ${\bf x}|\omega_0,\varsigma,\mu$ as
\begin{equation}
	p(x_1,\ldots,x_n|\omega_0,\varsigma,\mu) = p(x_1|\omega_0,\varsigma,\mu) \prod_{i=2}^n p(x_i|x_{i-1},\omega_0,\varsigma,\mu) \label{eq-oulike}.
\end{equation}
Under the assumption that the OU process is Gaussian and stationary, i.e., that the probability distribution of the process is not varying in time, the probability of the first sampled value is just the stationary distribution,
\begin{equation}
p(x_1|\omega_0,\mu,\varsigma) = \frac{1}{\sqrt{2\pi R_{OU}(0)}} \exp \left \{-\frac{1}{2}\frac{(x_1 - \mu)^2}{R_{OU}(0)} \right \} \label{eq-ou_sprob}.
\end{equation}
For a Gaussian OU process, the probability distribution of $x_i|x_{i-1}$ can be calculated from Equation (\ref{eq-ousol}) \citep[e.g.,][]{Brockwell2002a} as
\begin{align}
	\begin{split}
	& p(x_i|x_{i-1},\omega_0,\varsigma,\mu) = \\
	& \quad \frac{1}{\sqrt{2\pi R_{OU}(0)(1-\rho_i^2)}} \exp\left\{-\frac{1}{2}\frac{(\tilde{x}_i - \rho_i \tilde{x}_{i-1})^2}{R_{OU}(0) (1 - \rho_i^2)}\right\} \end{split} \label{eq-ou_cprob} \\
	& \quad \quad \quad \quad \quad \quad \quad \tilde{x}_i = x_i - \mu \label{eq-xtilde} \\
	& \quad \quad \quad \quad \quad \quad \quad \rho_i = e^{-\omega_0 (t_i - t_{i-1})} \label{eq-ou_rho}.
\end{align}
Equations (\ref{eq-oulike})--(\ref{eq-ou_rho}) allow us to compute the likelihood function of a sampled OU process, and thus perform statistical inference on the variability parameters. We note that above we have assumed that the sampled data points ${\bf x}$ are measured without error. KBS09 provide the likelihood function when the sampled data are contaminated by normally distributed measurement errors, while \citet{Kozowski2010a} provide an alternative algorithm for calculating the likelihood function. Below in \S~\ref{s-counts} we discuss the situation in which the sampled data points consist of a series of individual photon count measurements, as is the case for X-ray or $\gamma$-ray lightcurves.

\subsection{Superpositions of Ornstein-Uhlenbeck Processes}
\label{s-supou}

KSS11 introduced a superposition of OU processes as a model for the X-ray lightcurves generated by black holes. The sup-OU processes are able to model processes exhibiting "long-memory" and have been used in a variety of other disciplines \citep[e.g.,][]{Barndorff-Nielsen2001a}. Long-memory processes have a PSD that decays as $1 / \omega^{\alpha}$ with $0 < \alpha < 1$ as $\omega \rightarrow 0$, and therefore exhibit correlations that decay slowly with time. The PSDs of X-ray lightcurves for both AGN and galactic black holes in the soft state exhibit this type of behavior\footnote{The PSDs of galactic black holes exhibit more structure than a simple broken power-law form, with a sum of Lorentzian functions often providing a good fit. However, as we are primarily interested in lightcurves of counts the simple broken power-law model will be sufficient for modeling such low $S/N$ lightcurves.}; galactic black holes in the `hard-state' also have an additional low-frequency break in their PSD where the PSD flattens from a form with $\alpha \approx 1$ to one with $\alpha \approx 0$ toward the lowest frequencies sampled. In addition, KSS11 showed that the sup-OU process is a solution to the linear stochastic diffusion equation, providing an interpretation of the parameters within the context of accretion physics.

For a set of $M$ independent zero-mean OU processes $X_1(t), \ldots, X_M(t)$ with characteristic frequencies $\omega_1, \ldots, \omega_M$ and fixed RSVP $\varsigma^2$, we construct a sup-OU process as
\begin{equation}
	Y_M(t) = \mu + \sum_{j=1} c_j X_j(t), \label{eq-supou}
\end{equation}
where $c_1, \ldots, c_M$ are the mixing weights. The autocovariance and PSD of this process are, respectively,
\begin{align}
	R_{Y,M}(t) & = \frac{\varsigma^2}{2} \sum_{j=1}^M \frac{c_j^2}{\omega_j} e^{-\omega_j |t|} \label{eq-supou_autocov} \\
	P_{Y,M}(t) & = \frac{\varsigma^2}{2\pi} \sum_{j=1}^M \frac{c_j^2}{\omega_j^2 + \omega^2}. \label{eq-supou_psd}
\end{align}

The PSDs of X-ray lightcurves from AGN and lower $S/N$ lightcurves from galactic black holes can be well described as a doubly-broken power-law
\begin{equation}
	P(\omega) \propto \begin{cases}
		1 & \omega \ll \omega_L \\
		1 / \omega^{\alpha} & \omega_L \ll \omega \ll \omega_H \\
		1 / \omega^2 & \omega \gg \omega_H
	\end{cases} \label{eq-plaw_psd}
\end{equation}
for $0 < \alpha < 2$; typically $\alpha \approx 1$. The parameters $\omega_L$ and $\omega_H$ define the low and high break frequency, respectively, and correspond to time scales at which the variability properties of the accretion flow emission change. In order to model a PSD of the form of Equation (\ref{eq-plaw_psd}) KSS11 used the following sequence of characteristic frequencies and mixing weights:
\begin{align}
	\log \omega_j & = \log \omega_L + \frac{j-1}{M-1} \log (\frac{\omega_H}{\omega_L}), j = 1, \ldots, M \label{eq-supou_omegas} \\
	c_j & = \omega_j^{1 - \alpha / 2} \left( \sum_{j=1}^M \omega_j^{2-\alpha} \right)^{-1/2}. \label{eq-supou_weights}
\end{align}
Within this weighting scheme the $1 / \omega^{\alpha}$ part of the PSD is built up by the `knees' of the PSDs of the $M$ individual OU processes. In addition, under this weighting scheme the normalization of the PSD on frequencies $\omega \gg \omega_H$ is still proportional to $\varsigma^2$, so we will also refer to $\varsigma^2$ for the sup-OU process as the RSVP.

The sup-OU process is not Markovian, so the probability distribution for a sampled process $x_1, \ldots, x_n$ as a function of the parameters $\theta = (\mu, \varsigma, \omega_L, \omega_H, \alpha)$ is factored as
\begin{equation}
	p(x_1,\ldots,x_n|\theta) = p(x_1|\theta) \prod_{i=2}^n p(x_i|{\bf x}_{<i},\theta) \label{eq-supoulike}.
\end{equation}
Here, ${\bf x}_{<i} = [x_1, \ldots, x_{i-1}]^T$. Computation of Equation (\ref{eq-supoulike}) is more complicated than Equation (\ref{eq-oulike}) and is done using the Kalman recursions \citep[e.g., see][KSS11]{Brockwell2002a}. KSS11 derive an efficient method for calculating the likelihood function under normally distributed measurement errors, and Equation (\ref{eq-supoulike}) can be computed using their Equations (30)--(41) while setting the measurement error variances to zero.

\subsection{OU and sup-OU Processes for a Time Series of Counts}
\label{s-counts}

In the previous sections we have provided the mathematical background for the OU and sup-OU processes, as well as the likelihood function, which can be used to fit the variability parameters from a sampled version of these processes. However, in reality we do not observe a sampled version of the process (i.e., the true lightcurve values), and in practice we have to introduce a measurement process that connects sampled values of the true lightcurve to a measured lightcurve. When the lightcurve is contaminated with normally distributed measurement error, then the likelihood function for the measured lightcurve is given by KBS09 for the OU process model and by KSS11 for the sup-OU process model; we refer the reader to those papers for further details. However, for X-ray and $\gamma$-ray data it is often the case that one detects individual photons, and therefore the measurements consist of a time series of photon counts\footnote{This is a simplification. What are actually observed are the photon arrival times. The number of photons arriving within a given time bin are counted, creating the measured time series of counts.}. In this case the measurement process is a Poisson process, with the rate parameter controlled by the unknown true count rate within the source extraction region. This true count rate is a mixture of the source count rate and the background count rate. In this work we model the source count rate as a sup-OU process with weighting scheme given by Equation (\ref{eq-supou_weights}).

Denote the measured counts within the source extraction region in a time bin $t_i$ as $S_i$. The measured counts within the source extraction region follows a Poisson distribution with rate parameter given by the sum of the source and background count rates:
\begin{equation}
	p(S_i|s_i, b_i) = \frac{(\delta_i(s_i+b_i))^{S_i}}{S_i!}e^{-\delta_i(s_i+b_i)}. \label{eq-poisson}
\end{equation}
Here, $\delta_i, s_i,$ and $b_i$ are the exposure time, source count rate, and background count rate for the $i^{\rm th}$ time bin, respectively. Similarly, the measured counts within the background extraction region, $B_i$, also follows a Poisson distribution
\begin{equation}
	p(B_i|b_i) = \frac{(R_{\rm area} \delta_i b_i )^{B_i}}{B_i!}e^{-R_{\rm area} \delta_i b_i}, \label{eq-bpoisson}
\end{equation}
where $R_{\rm area}$ is the ratio of the area of the background extraction region to the source extraction region. Equations (\ref{eq-poisson}) and (\ref{eq-bpoisson}) provide the necessary equations that enable us to link the measured counts to the variability parameters.

Based on the Poisson measurement process, we have the following hierarchical model that relates the measured time series of counts to the sup-OU process parameters, $\theta$:
\begin{align}
	S_i|s_i, b_i & \sim {\rm Poisson}(\delta_i(s_i+b_i)), \label{eq-hmsource_counts} \\
	B_i|b_i & \sim {\rm Poisson}(R_{\rm area} \delta_i b_i), \label{eq-hmback_counts} \\
	\log {\bf s} | \mu,\varsigma,\omega_L,\omega_H,\alpha & \sim 
		{\rm sup\text{-}OU}(t_1,\ldots,t_n,\mu,\varsigma,\omega_L,\omega_H)  \label{eq-hmsource} \\
	b_i,\ldots,b_n|\theta_b & \sim p(b_1,\ldots,b_n|t_1,\ldots,t_n,\theta_b) \label{eq-hmback} \\
	\mu,\varsigma,\omega_L,\omega_H,\alpha,\theta_b & \sim p(\mu,\varsigma,\omega_L,\omega_H,\alpha,\theta_b) 
	\label{eq-hmprior}.
\end{align}
Here the notation $x \sim p(x)$ means that $x$ has the probability distribution $p(x)$, ${\rm Poisson}(\lambda)$ denotes a Poisson distribution with rate parameter $\lambda$, ${\rm sup\text{-}OU}(t_1,\ldots,t_n,\mu,\varsigma,\omega_L,\omega_H,\alpha)$ denotes a sup-OU process sampled at the time values $t_1,\ldots,t_n$, $p(b_1,\ldots,b_n|t_1,\ldots,t_n,\theta_b)$ denotes the probability distribution of the background count rates at the sampled times as a function of its parameters $\theta_b$, and $p(\mu,\varsigma,\omega_L,\omega_H,\alpha,\theta_b)$ denotes the prior on the parameters. We will later specify $p(b_1,\ldots,b_n|t_1,\ldots,t_n,\theta_b)$, and possible choices include stochastic processes such as  an OU or sup-OU process, or deterministic forms such as a constant, polynomial, or spline. We model the logarithm of the count rates as a sup-OU process instead of the actual count rates because the sup-OU process can be negative. Because we are assuming a Gaussian sup-OU process, the marginal distribution of the count rates assuming Equation (\ref{eq-hmsource}) will be a log-normal distribution, in agreement with what is observed for the X-ray binary Cyg X-1 and what is implied by the rms--flux relationship of X-ray binaries and AGN \citep{Uttley2005a}. Finally, we note that one can replace Equation (\ref{eq-hmsource}) with an OU process to obtain the equivalent results when using an OU process model.

In order to derive the likelihood function of the source and background counts as a function of the parameters, we start with the likelihood function for both the unknown count rates and the measured counts (i.e., the `complete' data likelihood function) and then marginalize over the unknown values of the count rates. From Equation (\ref{eq-hmsource_counts})--(\ref{eq-hmback}) we derive the complete data likelihood function by multiplying the conditional probability distributions together:
\begin{equation}
	\begin{split}
	& p({\bf S}, {\bf B}, {\bf s}, {\bf b}|\mu,\varsigma,\omega_L,\omega_H,\alpha,\theta_b) = \\
	& \quad \prod_{i=1}^{n} p(\log s_i|{\bf s}_{<i},\mu,\varsigma,\omega_L,\omega_H,\alpha)
	p(b_i|{\bf b}_{<i}, \theta_b) \\
	& \quad \times \frac{(R_{\rm area} \delta_i b_i)^{B_i} (\delta_i [s_i + b_i])^{S_i}}{B_i!S_i!}
	e^{-\delta_i(s_i + [1 + R_{\rm area}] b_i)}.
	\end{split}
	 \label{eq-complete_lik}
\end{equation}
Here, ${\bf S}, {\bf B}, {\bf s},$ and  ${\bf b}$ denote vectors containing the values for these quantities for the $n$ data points, and ${\bf s}_{<i}$ should be ignored for $i = 1$. Note that the background model need not depend on the earlier values, ${\bf b}_{<i}$, and if the background model is deterministic then $p(b_i|{\bf b}_{<i})$ is just a delta function centered at the model value of $b_i$. The measured data likelihood function is then found by integrating Equation (\ref{eq-complete_lik}) over the unknown values of ${\bf s}$ and ${\bf b}$:
\begin{equation}
	\begin{split}
	& p({\bf S}, {\bf B}|\mu,\varsigma,\omega_L,\omega_H,\theta_b) = \\
	& \quad \prod_{i=1}^n \int_{b_i=0}^{\infty} p(b_i|{\bf b}_{<i}, \theta_b) \frac{(R_{\rm area} \delta_i b_i)^{B_i}}{B_i!}
	e^{-R_{\rm area} \delta_i b_i} \\
	& \quad \quad \times \left \{ \int_{\log s_i=-\infty}^{\infty} p(\log s_i|{\bf s}_{<i},\mu,\varsigma,\omega_L,\omega_H) \right. \\
	& \quad \quad \quad \times \left. \frac{(\delta_i [s_i + b_i])^{S_i}}{S_i!} e^{-\delta_i (s_i + b_i)}\ d\log s_i \right \} db_i
	\end{split}
	\label{eq-measured_lik}
\end{equation}
A maximum-likelihood fit of the sup-OU parameters can be obtained by maximizing Equation (\ref{eq-measured_lik}) with respect to $(\mu,\varsigma,\omega_L,\omega_H,\alpha,\theta_b)$. Alternatively, one can obtain the posterior probability distribution of $(\mu,\varsigma,\omega_L,\omega_H,\alpha,\theta_b)$ given the measured counts by combining Equation (\ref{eq-measured_lik}) with the prior distribution used in Equation (\ref{eq-hmprior}).

\subsection{Practical Implementation of OU Process Models to Count Data}
\label{s-implementation}

The integral in Equation (\ref{eq-measured_lik}) is intractable and Equation (\ref{eq-measured_lik}) is not used in practice. Instead, the integral in Equation (\ref{eq-measured_lik}) is evaluated stochastically \citep[e.g.,][]{Jung2006a}, as in, for example, Monte Carlo integration. In Bayesian inference this is typically done by regarding the unknown source and background count rates, ${\bf s}$ and ${\bf b}$, as additional parameters to be estimated. This is often referred to as `data augmentation' \citep[e.g.,][]{Gelman2004a}, and this is the approach employed in this work.

The posterior probability distribution of the count rates and the variability parameters, given the measured counts, is proportional to Equation (\ref{eq-complete_lik}). One then uses an MCMC sampler to generate random draws of the count rates and parameters from their posterior. This has the advantage that we also obtain an estimate of the true count rate and its uncertainty for a source, which can be helpful for visualizing the lightcurve. In this work we employ an MCMC sampler based on an ancillarity-sufficiency interweaving strategy \citep[ASIS,][]{Yu2011a}. The basic idea behind this approach is to improve MCMC efficiency by updating the source or background count rates and their parameters under two different data augmentations.  A na\"{i}ve MCMC sampler under the parameterization described by Equations (\ref{eq-hmsource_counts})--(\ref{eq-hmprior}) would update the values of $s_i$ one at a time. This is very inefficient due to large correlations in the posterior of $s_1,\ldots,s_n$, and the addition of a second data augmentation significantly improves the efficiency of the MCMC sampler \citep{Yu2011a}. When updating the values of the count rates and parameters we use a Metropolis algorithm with normally distributed proposals centered at the current value.

Even with the ASIS steps, we still experienced slow convergence of our MCMC sampler. In order to improve convergence we introduced a few simplifications to our sampler. First, we fix the values of $\omega_L$ and $\alpha$. We do this because in this work we are primarily interested in estimating $\varsigma$, and because $\varsigma$ is proportional to the normalization of the PSD on frequencies $\omega \gg \omega_H$ the posterior distribution for $\varsigma$ is not sensitive to the values of $\omega_L$ and $\alpha$. Moreover, in general we do not expect to significantly sample the PSD at frequencies $\omega \ll \omega_H$ due to our use of X-ray lightcurves which in general are $\lesssim 100$ ksec. For comparison, the value of $\tau_H = 1 / \omega_H$ is expected to be $\sim$ ksec and $\sim$ days to weeks for a $10^7$ and $10^9 M_{\odot}$ black hole, respectively (e.g., KBS11). We fix the value of $\alpha$ to be equal to one, a typical value for AGN \citep[e.g.,][]{Uttley2005b}. We fix the value of $\omega_L$ to be the smaller of the inverse of ten times the length of the lightcurve or $10^{-5}\ {\rm ksec^{-1}}$. These values were chosen to correspond to sufficiently long time scales that we would not observe the low-frequency break in any of our lightcurves.

We model the background count rates, $b_1,\ldots,b_n$ as being generated by an OU process. This is probably not exactly true in practice, but the point of this model is to provide estimates of the background count rates which are smoothed versions of the measured counts in each bin. We estimate the background rates by first running our MCMC sampler only on the counts extracted from the background region, using an OU process model instead of a sup-OU process model. In this step the $b_1,\ldots,b_n$ play the role of the $s_1,\ldots,s_n$ in Equations (\ref{eq-hmsource_counts})--(\ref{eq-hmprior}) and the background counts and count rates are set to zero. Our MCMC sampler then returns a sample of $b_1,\ldots,b_n$ from its probability distribution given the measured counts extracted from the background region, under the prior assumption that it follows an OU process. We then estimate the background count rates as the median values of $b_i$ obtained in this manner, providing smoothed estimates of the background counts. These values of $b_1,\ldots,b_n$ are then held fixed when running our MCMC sampler on the counts extracted from the source region. The accuracy of this approximation improves with the $S/N$ of the background lightcurve, so it is best to use a much larger background extraction region compared to the source extraction region. We have performed simulations to confirm that our approach for handling the background does not introduce significant biases on the estimation of the source parameters, so long as the background count rate do not dominate the source count rate. If the background dominates in certain time intervals, then it is best to remove these intervals.

We assume a uniform prior on the parameters $(\mu, \varsigma, \log \omega_H)$. The mean of the logarithm of the count rate is constrained to be $\ln 10^{-6} < \mu < \ln 10^6$ and the location of the high frequency break is constrained to be $\omega_L < \omega_H < 1 / \min(t_{i+1} - t_i)$.

\subsection{Illustration with Simulated Data}
\label{s-simulation}

\begin{figure}
	\includegraphics[scale=0.5,angle=00]{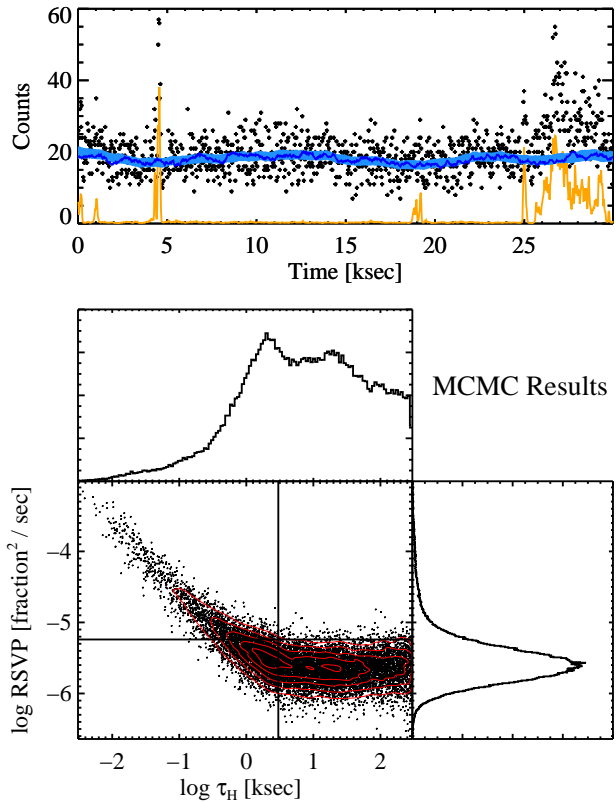}
	\caption{Illustration of our new statistical technique on a simulated lightcurve of counts. The top panel shows the measured counts (data points), the true count rate for the simulated lightcurve (darker blue line), the background rate (orange line). The lighter blue shaded region contains $90\%$ of the posterior probability on the count rate derived from our Bayesian approach. The bottom panels show the posterior probability distribution (histograms and red contours) for the high-frequency break time scale $\tau_H$ and the RSVP $\varsigma^2$, as estimated from the MCMC sampler output (data points). The two straight lines mark the true values used to simulate the lightcurve. The true values of the lightcurve, $\tau_H$, and $\varsigma^2$ are recovered from the counts.
	\label{f-illust}}
\end{figure}

In order to illustrate our method, we simulated a 30 ksec lightcurve assuming a sup-OU process with a value of $\tau_H = 3$ ksec, $\varsigma = 2.4 \times 10^{-3}$ cnt / sec$^{1/2}$, and a mean count rate of 0.45 counts / sec. We used a low-frequency break corresponding to a time scale a factor of 100 longer than the length the lightcurve to ensure that we were always on the $1 / f$ or $1 / f^2$ part of the PSD, as is typical for AGN. Our background rate was chosen to be a resampled and wavelet-denoised version of the background from the XMM observation of PG 2130+099. We chose the background for this object because it contains regions of flaring activity, and thus should provide a good test of our technique. We then simulated source and background counts using a bin size of 30 sec, and assumed that the background extraction region was a factor of seven larger than the source extraction region. We applied our MCMC sampler to this simulated lightcurve, running four independent chains with a total of $1.5 \times 10^5$ iterations including $5 \times 10^4$ iterations of burn-in. The results are shown in Figure \ref{f-illust}. Our technique is able to recover the simulated lightcurve as well as the variability parameters within the uncertainties from the count data. 

The bumpiness in the posterior distribution for $\log \tau_H$ is Monte Carlo error due to slow convergence and poor mixing of the chains. If we had run our MCMC sampler for longer the posterior would be smoothed out. We also observed this for the $\log \tau_H$ posteriors obtained for the AGN in our sample. We do not consider this a concern because we are primarily interested in the posterior for $\log \varsigma^2$, and in particular lower order summaries of the posterior such as the median and standard deviation, as well as upper limits in the case of no detection. Low order summaries converge must faster in an MCMC sampler than, say, the tails of the distribution. 

\section{Trends Between X-ray Variability, Optical Variability, and Black Hole Mass}
\label{s-trends}

The X-ray and optical RSVP and their uncertainties are given in Table \ref{t-rsvp}.

We applied our new method for deriving the sup-OU parameters from a lightcurve of counts to those sources in Table \ref{t-sample} that had X-ray lightcurves. For the sources already analyzed by KSS11 we used the values of $\varsigma$ derived by them, as those sources are bright enough that the X-ray counts are well approximated as coming from a normal distribution, and because their values are derived from a combination of XMM and RXTE lightcurves. For each source not in the KSS11 sample we ran four parallel chains of our MCMC sampler for $3 \times 10^5$ iterations and discarded the first $10^5$ iterations as burn-in. We also used an MCMC sampler based on the Gaussian approximation method of KSS11 as a check for consistency; for all but the faintest sources the two methods produced similar posteriors for $\varsigma$. 

Unfortunately, in spite of the large number of iterations used when running our MCMC sampler, we still had poor convergence of $\varsigma$ when directly analyzing the count data for Mrk 79, NGC 5548, 3C 390.3, Mrk 50, and 3C 273. The MCMC sampler of KSS11 based on the Gaussian approximation converges much faster, so we used the method of KSS11 on these four sources. These sources are all bright enough to justify the Gaussian approximation, having count rates of $\gtrsim 1$ counts / sec. In addition, we were only able to obtain an upper limit on $\varsigma$ for PG 0052+251, PG 0804+761, PG 0953+414, PG 1229+204, PG 1307+085, Mrk 279, and PG 1411+442.

The \emph{Swift}-XRT lightcurve for PG 1426+015 exhibits strong variability, and the amplitude of variability is unusual for a source with a 
$M_{BH} \sim 10^9 M_{\odot}$ black hole. For this source we obtained a value of $\log \varsigma^2_X = -5.49 \pm 0.37\ {\rm fraction^2 / sec}$ 
and $\tau_H \lesssim 1$ day. This is considerably more variable than expected for such a massive black hole. Moreover, the timescale 
corresponding to the high-frequency break is much shorter than the value of $\tau_H \sim 30$--$100$ days expected for $M_{BH} \sim 10^9 
M_{\odot}$ \citep[e.g.,][]{McHardy2006a,Kelly2011a,Gonzalez-Martin2012a}. For example, between the 20--26 day markers in Figure \ref{f-pg1426_lightcurve} the count rate triples over a couple of days, and then drops to half this value a couple of days later. This strong flare 
corresponds to the count rate obtained from segment 12; such a strong flare is unexpected for this type of object, a radio-quiet quasar with $M_{BH} \sim 10^9 M_{\odot}$. The count rate of this segment of the 
lightcurve differs from that expected based on the best-fit sup-OU process by $\sim 2.7\sigma$, suggesting that this data point is an outlier. The primary reason for the large count rate is the large correction factor for this segment calculated by the 
\texttt{xrtlccorr} routine, probably due to the fact that the source fell further off-axis for this observation segment compared to the others; the count rate 
obtained before performing this correction was not particularly unusual. Whether this data point is an outlier due to systematics or real flaring 
activity, this data point is inconsistent with being generated from a sup-OU process and is indicative of coming from a different variability 
process. Because of this we removed this data point. After removing this data point we find more reasonable constraints on the variability 
timescale of $\tau_H \lesssim 15$ days, although values closer to $\sim 1$ day are still preferred. In addition, the value of X-ray RSVP is 
reduced but still high at $\log \varsigma^2_X = -6.16 \pm 0.33$. The values we report in Table \ref{t-rsvp} for PG 1426+015 are 
obtained after removing segment 12 of the lightcurve.

For the optical data we used the MCMC sampler based on the CAR(1) model described by KBS09. We subtracted the host galaxy flux from the optical lightcurves using the values from \citet{Bentz2009a,Bentz2013a}. We then ran our MCMC sampler on the logarithm of the flux values to ensure that $\varsigma^2_{\rm opt}$ is in units of fraction$^2$ / sec.

 \begin{deluxetable}{lcccc}
\tabletypesize{\scriptsize}
\tablewidth{0pt}
\tablecaption{X-ray and Optical Rate of Stochastic Variability Power \label{t-rsvp}}
\tablehead{
Name
& $\log \varsigma_X^2$ 
& $\log \varsigma^2_X$ Error\tablenotemark{a} 
& $\log \varsigma^2_{\rm opt}$ 
& $\log \varsigma^2_{\rm opt}$ Error
}
\startdata
3C 120      & -7.21 & 0.16 & -8.25 & 0.11 \\
3C 390.3    & -7.79 & 0.28 & -7.83 & 0.14 \\
Ark 120     & -7.15 & 0.13 & -7.62 & 0.14 \\
Fairall 9   & -7.84 & 0.13 & -7.75 & 0.17 \\
MRK 766     & -5.15 & 0.05 & $\ldots$ & $\ldots$ \\
Mrk 110     & -6.69 & 0.21 & -7.7 & 0.09 \\
Mrk 279     & -6.22 & 0.0 & -7.94 & 0.13 \\
Mrk 290     & $\ldots$ & $\ldots$ & -8.23 & 0.09 \\
Mrk 335     & -5.1 & 0.05 & -8.47 & 0.12 \\
Mrk 50      & -6.32 & 0.32 & $\ldots$ & $\ldots$ \\
Mrk 509     & -7.03 & 0.05 & $\ldots$ & $\ldots$ \\
Mrk 590     & -6.65 & 0.42 & -7.62 & 0.09 \\
Mrk 6       & -6.67 & 0.25 & $\ldots$ & $\ldots$ \\
Mrk 766     & $\ldots$ & $\ldots$ & -7.81 & 0.13 \\
Mrk 79      & -6.6 & 0.18 & -8.3 & 0.09 \\
Mrk 817     & -5.54 & 0.38 & -8.86 & 0.1 \\
NGC 3227    & -6.05 & 0.06 & -7.37 & 0.09 \\
NGC 3516    & -5.9 & 0.05 & -7.34 & 0.1 \\
NGC 3783    & -6.81 & 0.05 & -7.53 & 0.12 \\
NGC 4051    & -4.56 & 0.04 & -7.52 & 0.1 \\
NGC 4151    & -6.47 & 0.07 & -6.81 & 0.14 \\
NGC 4395    & -3.63 & 0.08 & $\ldots$ & $\ldots$ \\
NGC 4593    & -5.62 & 0.23 & -7.21 & 0.22 \\
NGC 5548    & -6.75 & 0.12 & -7.66 & 0.03 \\
NGC 6814    & -5.42 & 0.13 & -7.48 & 0.11 \\
NGC 7469    & -6.21 & 0.08 & -7.78 & 0.16 \\
PG 0026+129 & $\ldots$ & $\ldots$ & -8.31 & 0.08 \\
PG 0052+251 & -4.93 & 0.0 & -8.1 & 0.09 \\
PG 0804+761 & -6.8 & 0.56 & -8.77 & 0.08 \\
PG 0953+414 & -4.67 & 0.0 & -8.95 & 0.1 \\
PG 1226+023 & -7.88 & 0.13 & -9.09 & 0.1 \\
PG 1229+204 & -5.16 & 0.0 & -7.7 & 0.11 \\
PG 1307+085 & -4.46 & 0.0 & -8.44 & 0.12 \\
PG 1411+442 & -3.58 & 0.0 & -7.85 & 0.12 \\
PG 1426+015 & -6.16 & 0.33 & -8.52 & 0.1 \\
PG 1613+658 & $\ldots$ & $\ldots$ & -8.62 & 0.08 \\
PG 1617+175 & $\ldots$ & $\ldots$ & -8.36 & 0.09 \\
PG 1700+518 & $\ldots$ & $\ldots$ & -9.32 & 0.15 \\
PG 2130+099 & -6.16 & 0.3 & -8.65 & 0.08 \\
Zw 229-015  & -5.41 & 0.09 & $\ldots$ & $\ldots$ \\
\enddata

\tablecomments{The units of both $\varsigma_X^2$ and $\varsigma^2_{\rm opt}$ are fractional variance / sec.}
\tablenotetext{a}{The values of $\varsigma_X^2$ with an error of zero are upper limits.}

\end{deluxetable}

\subsection{X-ray Variability as a Black Hole Mass Estimator}
\label{s-xray_bhole}

\begin{figure}
	\includegraphics[scale=0.5,angle=0]{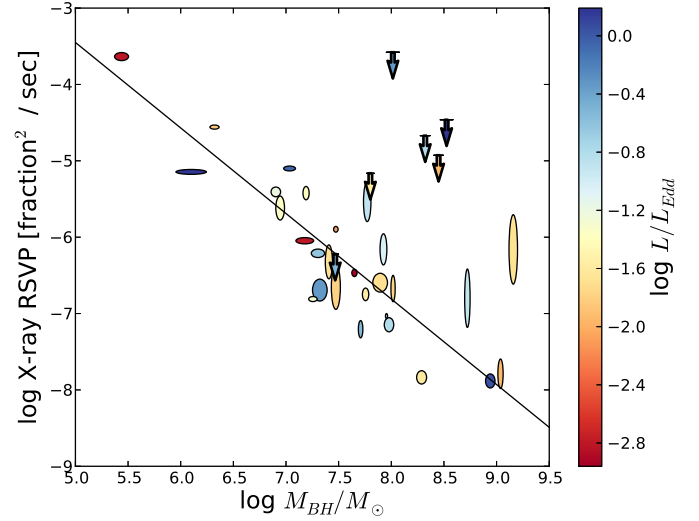}
	\caption{X-ray rate of stochastic variability power, $\varsigma_X^2$, as a function of black hole mass. The sizes of the ellipses denote the size of the error bars, and their colors denote the Eddington ratios. The arrows denote upper limits, defined to be the $99^{\rm th}$ percentile of the posterior probabilities for $\varsigma_X^2$.
	\label{f-mbh_rsvpx}}
\end{figure}

In Figure \ref{f-mbh_rsvpx} we compare the X-ray RSVP, $\varsigma^2_X$, with black hole mass. A strong anti-correlation is apparent. To quantify this anti-correlation (and all other linear fits in this work), we used the Bayesian linear regression method of \citet{Kelly2007a} which accounts for the uncertainty in both $\varsigma$ and $M_{BH}$, as well as the upper limits on $\varsigma$:
\begin{equation}
	\begin{split}
	& \log \frac{\varsigma^2_X}{\rm fraction^2\ sec^{-1}} = \\
	& \quad \quad (-6.81 \pm 0.13) - (1.12 \pm 0.13)
	\log \frac{M_{BH}}{10^8 M_{\odot}}.
	\end{split}
	\label{eq-mbh_rsvpx}
\end{equation}
The intrinsic scatter about this relationship is estimated to be ${\rm std}(\log \varsigma^2_X|M_{BH}) = 0.46 \pm 0.10$, where the notation ${\rm std}(x|y)$ denotes the standard deviation in $x$ at fixed $y$. The small intrinsic scatter about the relationship implies that $\approx 72\%$ of the variations in $\varsigma^2_X$ for our sample are driven by variations in $M_{BH}$. Our result is consistent with other recent work based on the X-ray excess variance \citep[e.g.,][]{Zhou2010a,Ponti2012a} or normalization of the high-frequency X-ray PSD \citep{Gierlinski2008a,McHardy2013a}, the results from which imply $\varsigma^2_X \propto 1 / M_{BH}$.  We did not find a statistically significant additional dependence on the bolometeric luminosity, suggesting that the dependence of $\varsigma^2_X$ on the Eddington ratio at fixed $M_{BH}$ is weak or nonexistent. Moreover, at fixed $M_{BH}$ we did not find any statistically significant evidence for a dependence of $\varsigma^2_X$ on optical luminosity, redshift or radio-loudness, where the radio flux values came from NED\footnote{\url{http://ned.ipac.caltech.edu}}.

\begin{figure}
	\includegraphics[scale=0.5,angle=90]{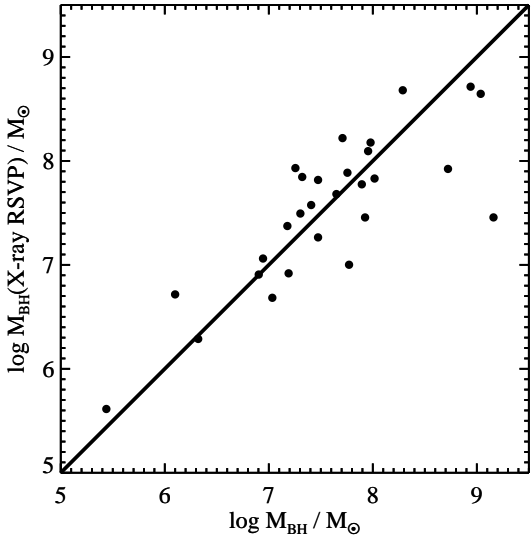}
	\caption{Comparison of the $\varsigma_X$--based mass estimates against those from the literature (mostly from reverberation mapping). The solid line marks the identity relationship; this is another way of representing the trend in Figure \ref{f-mbh_rsvpx}, but better highlights the use of $\varsigma^2_X$ as a mass estimator. The linear relationship defined by Equation (\ref{eq-mbh_rsvp}) provides a good fit, and there is no evidence for non-linearity in $M_{BH}$ as a function of $\varsigma_X$.}
	\label{f-mass_estimates}
\end{figure}

The tight anti-correlation between the X-ray RSVP and black hole mass suggests that $\varsigma^2_X$ may be used as a black hole mass estimator. Fitting $M_{BH}$ as a function of $\varsigma^2_X$ we find
\begin{equation}
	\begin{split}
	& \log \frac{M_{BH}}{M_{\odot}} = (8.07 \pm 0.10) \\ 
	& \quad - (0.73 \pm 0.09) \log \frac{\varsigma^2_X}{10^{-7} {\rm fraction^2\ sec^{-1}}}.
	\end{split}
	\label{eq-mbh_rsvp}
\end{equation}
The standard deviation in $M_{BH}$ at fixed $\varsigma$ is ${\rm std}(\log M_{BH}|\varsigma_X) = 0.38 \pm 0.08$. This result implies that mass estimates derived from the X-ray RSVP may be obtained with precision comparable to those derived from the rest-frame optical/UV spectra \citep{Vestergaard2006a}, and from the $M_{BH}$--$\sigma_*$ relationship \citep[e.g.,][]{McConnell2013a}. Figure \ref{f-mass_estimates} compares the mass estimates derived from $\varsigma_X^2$ with the literature values.

\begin{figure*}
	\includegraphics[scale=0.9,angle=0.0]{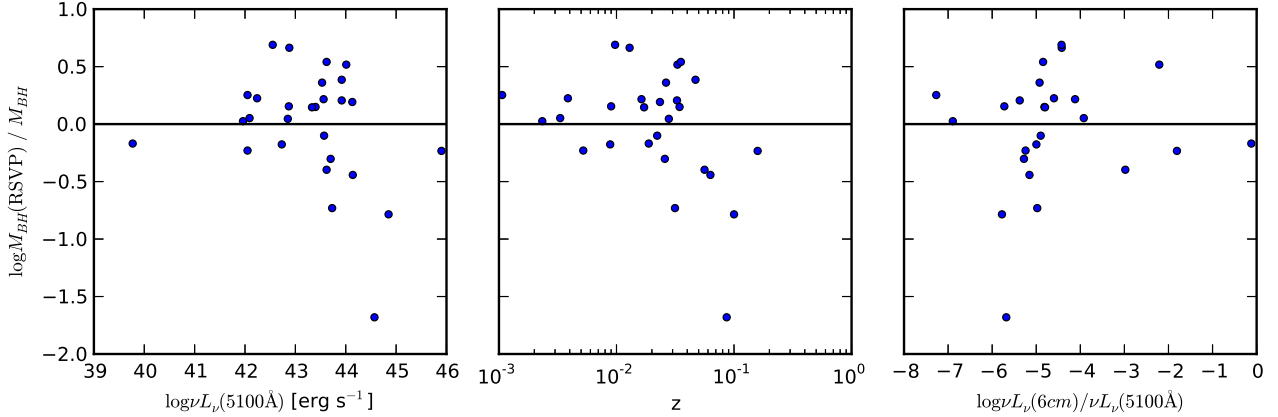}
	\caption{Comparison of the residuals in the X-ray RSVP-based $M_{BH}$ estimates with luminosity, redshift, and radio-loudness. There is no statistically significant evidence that any of these quantities acts as a `third parameter' in the $\varsigma_X$--$M_{BH}$ relationship, as the residuals in the $\varsigma_X$-based mass estimates are uncorrelated with these quantities.}
	\label{f-resid_trends}
\end{figure*}

In Figure \ref{f-resid_trends} we compare the residuals of the $\varsigma_X$-based mass estimates with optical luminosity, redshift, and radio-loudness. We did not find statistically significant evidence for any dependence of $M_{BH}|\varsigma_X$ on optical luminosity, redshift, or radio-loudness.

\subsection{Comparison between Optical Variability, X-ray Variability, and Black Hole Mass}
\label{optical_bhole}

\begin{figure}
	\includegraphics[scale=0.5,angle=0]{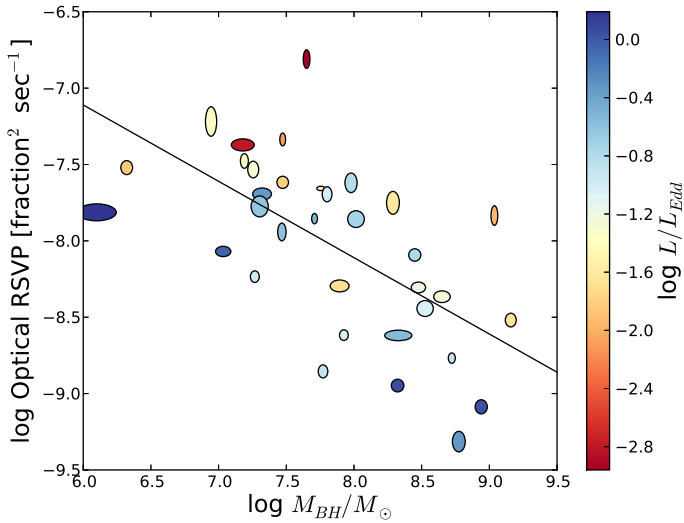}
	\caption{Optical rate of stochastic variability power, $\varsigma^2_{\rm opt}$, as a function of black hole mass. The symbols are the same as for Figure \ref{f-mbh_rsvpx}. The strong correlation in the optical RSVP with luminosity is manifested in the apparent correlation between $\varsigma^2_{\rm opt}$ and $L / L_{Edd}$ at fixed $M_{BH}$.
	\label{f-mbh_rsvpo}}
\end{figure}

In order to search for trends between the optical variability and $M_{BH}$, we performed the same analysis as in \S~\ref{s-xray_bhole}. In Figure \ref{f-mbh_rsvpo} we compare the optical RSVP, $\varsigma_{\rm opt}^2$, as a function of $M_{BH}$. There is a statistically significant anti-correlation between the optical RSVP and $M_{BH}$, which we quantify as
\begin{equation}
	\begin{split}
	& \log \frac{\varsigma^2_{\rm opt}}{\rm fraction^2\ sec^{-1}} = \\
	& \quad (-8.11 \pm 0.09) - (0.50 \pm 0.13) \log\left(\frac{M_{BH}}{10^8 M_{\odot}}\right).
	\end{split}
	\label{eq-opt_rsvp_mbh}
\end{equation}
The intrinsic scatter in $\varsigma^2_{\rm opt}$ at fixed $M_{BH}$ is ${\rm std}(\log \varsigma_{\rm opt}^2|M_{BH}) = 0.47 \pm 0.07$. This trend is not as strong as that in the X-rays. Moreover, performing a linear regression on the reverse relationship we find that the scatter in $M_{BH}$ at fixed optical RSVP is ${\rm std}(\log M_{BH}|\varsigma_{\rm opt}) = 0.59 \pm 0.09$, implying that mass estimates derived from this scaling relationship will not be as precise as those obtained from the other scaling relationships.

\begin{figure}
	\includegraphics[scale=0.4,angle=0]{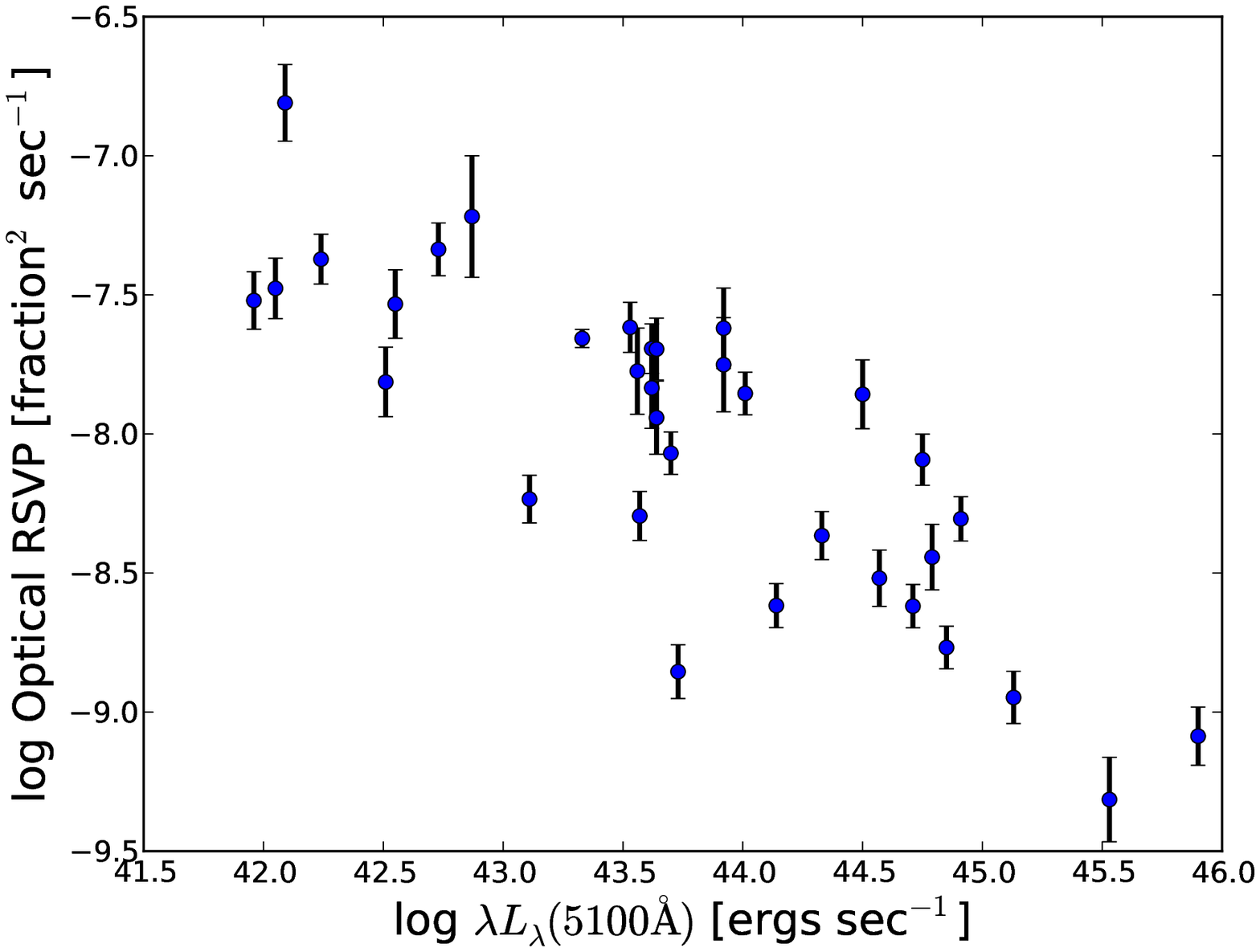}
	\includegraphics[scale=0.4,angle=0]{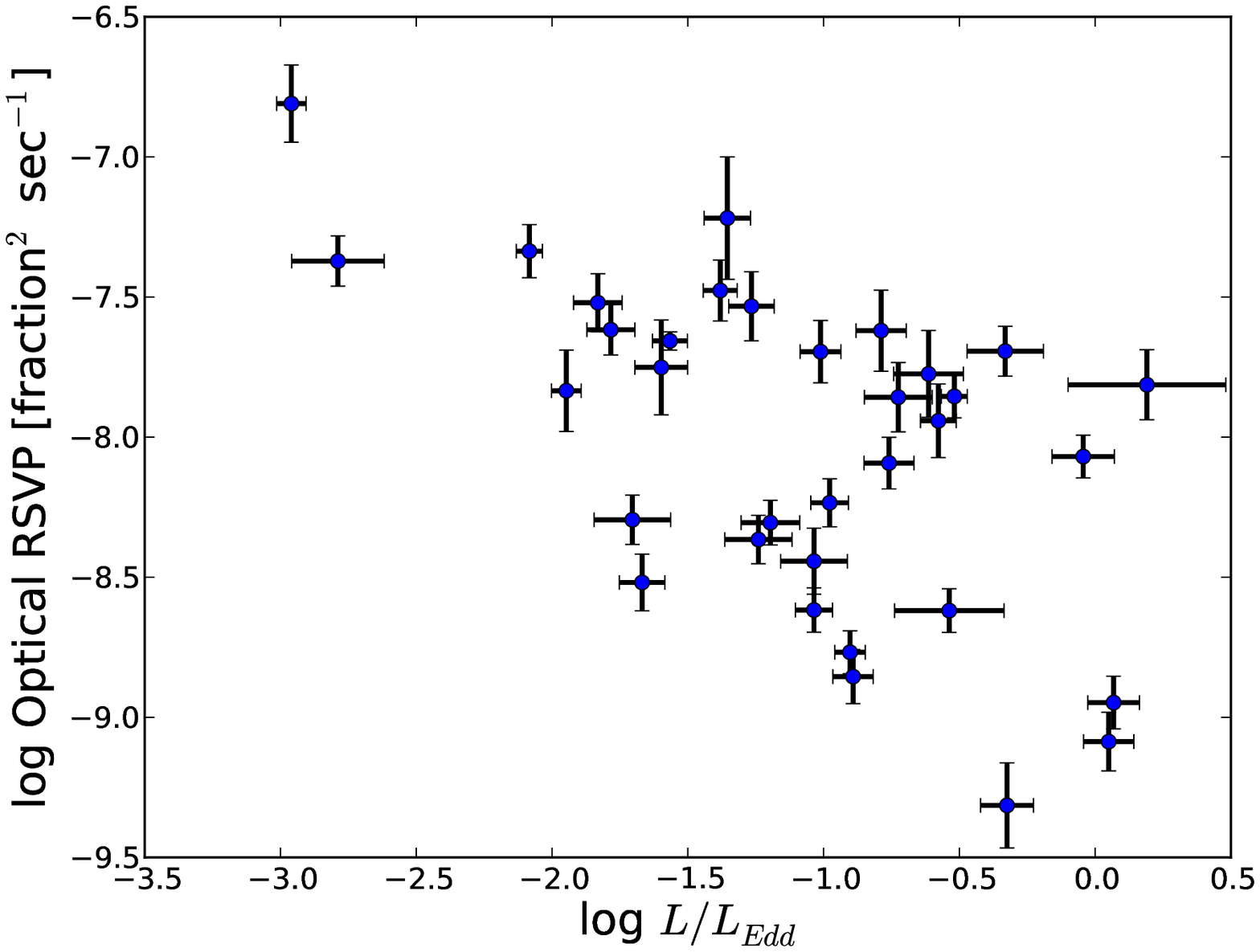}
	\caption{Optical rate of stochastic variability power, $\varsigma^2_{\rm opt}$, as a function of optical luminosity (top) and Eddington ratio (bottom). As with $M_{BH}$, there is a significant anti-correlation between these quantities, with brighter AGN becoming less variable on short time scales. 
	\label{f-lum_rsvpo}}
\end{figure}

Interestingly, although the optical RSVP does not show as tight of a correlation with $M_{BH}$ as does $\varsigma_X^2$, it is tightly anti-correlated with the luminosity. This correlation is implied by the observed increase in the optical RSVP with Eddington ratio at fixed $M_{BH}$ that is apparent in Figure \ref{f-mbh_rsvpo}. In Figure \ref{f-lum_rsvpo} we show $\varsigma^2_{\rm opt}$ as a function of the optical luminosity and $L / L_{Edd}$. The optical RSVP is most tightly correlated with optical luminosity:
\begin{equation}
	\begin{split}
	& \log \frac{\varsigma^2_{\rm opt}}{\rm fraction^2\ sec^{-1}} = \\
	& \quad (-8.14 \pm 0.06) - (0.47 \pm 0.06) \log \left(\frac{\lambda L_{\lambda}(5100\AA}{10^{44}{\rm\ erg\ s^{-1}}}\right). 
	\end{split}
	\label{eq-opt_rsvp_lbol}
\end{equation}
The intrinsic scatter about Equation (\ref{eq-opt_rsvp_lbol}) is ${\rm std}(\log \varsigma^2_{\rm opt}|L_{5100}) = 0.33 \pm 0.05$. The slope and scatter of the trend with Eddington ratio is very similar to that for $M_{BH}$. We did not find any statistically significant evidence for an additional dependence of optical RSVP on black hole mass at fixed optical luminosity. 

\begin{figure}
	\includegraphics[scale=0.33,angle=90]{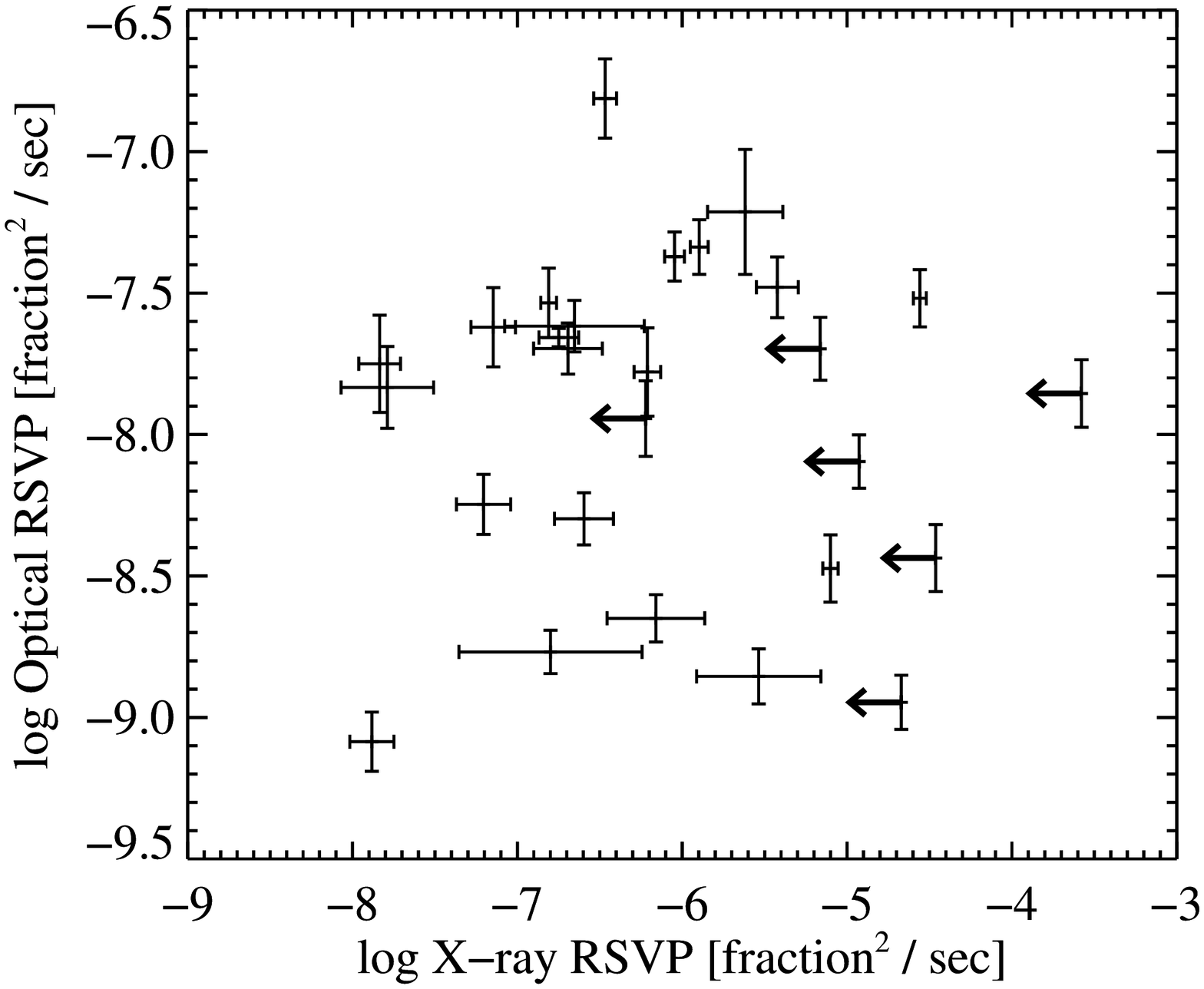}
	\caption{Comparison between the X-ray and optical RSVP. There is no evidence that the optical and X-ray variability amplitudes are correlated on time scales short compared to their respective break frequencies.
	\label{f-rsvp_compare}}
\end{figure}

In Figure \ref{f-rsvp_compare} we compare the optical RSVP against the X-ray RSVP. Despite the fact that both the optical and X-ray RSVP exhibit strong anti-correlations with some combination of $M_{BH}, L,$ and $L / L_{Edd}$, they are uncorrelated. This therefore implies that on time scales short compared to their respective break frequency, AGN that exhibit strong variable at, say, x-ray wavelengths will not necessarily exhibit strong variability at optical wavelengths. In addition there is a much larger spread in $\varsigma^2_X$ than $\varsigma^2_{\rm opt}$, implying that there is considerably more variety in the amplitude of X-ray variability than optical among AGN, at least over the time scale probed by this study.

\section{DISCUSSION}
\label{s-discussion}

\subsection{Comparison with Previous Work}
\label{s-comparison}

Previous work on comparing the amplitude of high-frequency variability with $M_{BH}$ has primarily focused on the X-ray excess variance, $\sigma^2_{\rm rms}$. The excess variance of a lightcurve on time scales $t_1 < \Delta t < t_2$ is given by the integral of the power spectrum over the corresponding frequency interval; for regularly sampled lightcurves $t_2$ corresponds to the length of the lightcurve and $t_1$ corresponds to the sampling interval. Neglecting sampling effects, when the length of the lightcurve is $t_2 \ll \tau_H$, then the excess variance and RSVP are proportional:
\begin{equation}
	\sigma^2_{\rm rms} \propto \varsigma^2_X (t_2 - t_1).
	\label{eq-rsvp_excess}
\end{equation}
However, for a real lightcurve the sampling pattern introduces a more complicated relationship between excess variance and $\varsigma^2_X$ \citep[e.g.,][]{Pessah2007a}, and Equation (\ref{eq-rsvp_excess}) only holds approximately.

Many authors have found that the scaling between X-ray excess variance, $\sigma^2_{\rm rms}$, and $M_{BH}$ is consistent with $\sigma^2_{\rm rms} \propto 1 / M_{BH}$ \citep[e.g,][]{Bian2003a,Nikolajuk2004a,Zhou2010a}. The previous works that are most similar to our analysis are the recent papers by \citet{Zhou2010a}, KSS11, \citet{Ponti2012a}, and \citet{McHardy2013a}. \citet{Zhou2010a} found $\sigma_{\rm rms}^2 \propto M_{BH}^{-1.00 \pm 0.10}$ based on a sample of 21 AGN with $M_{BH}$ from reverberation mapping and excess variance measurements taken from \citet{ONeill2005a}; \citet{ONeill2005a} calculated $\sigma^2_{\rm rms}$ from ASCA lightcurves having at least one X-ray observation with a duration of at least 40 ksec. These authors also estimate the intrinsic dispersion in $M_{BH}$ at fixed $\sigma^2_{\rm rms}$ to be ${\rm std}(\log M_{BH}|\sigma^2_{\rm rms}) = 0.20$ dex. \citet{Ponti2012a} calculated the excess variance from XMM observations for 32 AGN with $M_{BH}$ primarily from reverberation mapping. They calculated the excess variance from lightcurves of length 10, 20, 40, and 80 ksec, when available. Using the 10 ksec lightcurves, they find $\sigma^2_{\rm rms} \propto M_{BH}^{-1.21 \pm 0.1}$, and that the intrinsic scatter in the excess variance is ${\rm std}(\log \sigma^2_{\rm rms}|M_{BH}) \approx 0.45$ dex. \citet{McHardy2013a} compared $M_{BH}$ for 10 AGN with the amplitude of the power spectrum at $2 \times 10^{-4}$, normalized to the square of the source count rate, as calculated from EXOSAT observations longer than 20 ksec by \citet{Green1993a}; they refer to this quantity as the normalized variability amplitude, NVA. The mass estimates used in the \citet{McHardy2013a} study are primarily from reverberation mapping or dynamical modeling, with a couple estimated from the host galaxy $\sigma_*$. \citet{McHardy2013a} find ${\rm NVA}^2 \propto M_{BH}^{-1.08 \pm 0.08}$. Using a sample of 10 AGN, KSS11 found $\varsigma_X^2 \propto M_{BH}^{-1.58 \pm 0.44}$ based on a combination of RXTE and XMM ligthcurves, where the $M_{BH}$ estimates were from a combination of reverberation mapping, dynamical modeling, and scaling relationships. All of this previous work is consistent with our finding that $\varsigma^2_X \propto M_{BH}^{-1.12 \pm 0.13}$ and ${\rm std}(\log \varsigma^2_X|M_{BH}) = 0.46 \pm 0.10$ dex, especially considering the differences in samples, and that the excess variance can become a biased estimate of the normalization of the high-frequency PSD if the lightcurve is longer than $\tau_H$.

There has been little previous work comparing the amplitude of high-frequency optical variability and $M_{BH}$. Instead, most authors have focused on the total variability on longer time scales \citep[e.g.,][]{Wold2007a,MacLeod2010a,Zuo2012a}. KBS09 found $\varsigma^2_{\rm opt} \propto M_{BH}^{-0.52 \pm 0.08}$ from a heterogeneous sample of optical lightcurves for 71 AGN with $M_{BH}$ estimated primarily from the broad emission lines. Mass estimates from reverberation mapping were used for 20 of the AGN in the KBS09 study, but KBS09 did not remove the contribution from the host galaxy to the optical flux. In spite of these differences, their result is in excellent agreement with our result that $\varsigma^2_{\rm opt} \propto M_{BH}^{-0.50 \pm 0.13}$. However, while KBS09 do find a statistically significant but weak trend with luminosity, they were not able to conclude whether luminosity or $M_{BH}$ was the dominant correlation. This is in contrast to the result obtained here, where we find a strong trend with luminosity that dominates over the $M_{BH}$ correlation. This difference is probably due to the fact that in this work we subtracted the host galaxy contribution to the optical lightcurve before computing the optical RSVP.

\subsection{Implications for Black Hole Mass Estimation}
\label{s-mass_estimation}

The tight anti-correlation between $M_{BH}$ and the short time scale X-ray variability implies that the X-ray variability provides a means of estimating $M_{BH}$ that is competitive with other scaling relationships. The scatter in this relationship implies that $M_{BH}$ estimates derived from the X-ray RSVP have a precision of $\approx 0.38 \pm 0.08$ dex with respect to the $M_{BH}$ estimates derived in the literature. This is similar to $M_{BH}$ estimates obtained through the $M_{BH}$--$\sigma_*$ relationship for elliptical galaxies \citep[$\approx 0.34$ dex,][]{McConnell2013a} from the AGN broad emission lines \citep[$\sim 0.4$ dex,][]{Vestergaard2006a}. 

We also investigated the use of the optical RSVP as a mass estimator. The optical variability scaling relationship shows considerably more scatter, implying mass estimates with $\sim 0.6$ dex precision relative to the literature values of $M_{BH}$. Therefore it is unlikely that $\varsigma^2_{\rm opt}$-based mass estimates will be preferable to the other scaling relationships. However, it may still be possible to use the optical RSVP to obtain reasonable constraints on the demographics of $M_{BH}$ for a large sample of AGN optical lightcurves, such as would be obtained by, for example, LSST. This would be relevant for surveys that obtain optical lightcurves for large numbers of AGN, but lack the X-ray or spectroscopic observations needed to obtain more precise mass estimates from X-ray variability or the broad emission lines. In addition, the correlation of the residuals in the $\varsigma^2_{\rm opt}$--$M_{BH}$ relationship implies that the combination of optical RSVP and optical luminosity would provide more precise mass estimates compared to simply using $\varsigma_{\rm opt}^2$. Unfortunately, in order to accurately estimate the scatter in $M_{BH}$ at fixed $\varsigma_{\rm opt}^2$ and optical luminosity, and thus the precision in the mass estimates, it is necessary to obtain an unbiased sample with respect to the AGN Eddington ratio distribution. The sample analyzed in this work is heterogeneously selected, and thus unlikely represents the Eddington ratio distribution of the general AGN population. Further work using a sample that is unbiased with respect to Eddington ratio is needed in order to calibrate $M_{BH}$ estimates based on optical variability and luminosity.

\subsubsection{Sources of Systematic Uncertainty in the $M_{BH}$--$\varsigma_X^2$ Scaling Relationship}
\label{s-systematics}

We did not find any significant trends in the X-ray RSVP derived mass estimates with luminosity, redshift, or radio-loudness; the independence of the residuals with radio-loudness implies that the scaling relationship is not strongly affected by the presence of a jet. Previous work has found that the scaling relationship between the X-ray excess variance on short times scales does not exhibit a strong dependence on the spectral range used to compute the X-ray variance over the range 0.3--10 keV \citep[e.g.,][]{Ponti2012a}. We therefore do not expect a strong systematic effect with redshift so long as the X-rays detected from an AGN remain predominantly within this spectral range in the source's rest-frame. Because the number of photons / keV decays as $n(E) \propto E^{-\Gamma_X}$ for AGN, with $\Gamma_X \sim 2$ \citep[e.g.,][]{Young2009a}, most of the photons detected from an AGN will have rest-frame energies of $< 10$ keV even out to $z \sim 4$. 

It is also unclear if the $M_{BH}$--$\varsigma^2_X$ scaling relationship holds for obscured AGN, or for those radiating at $L / L_{Edd} \lesssim 10^{-2}$--$10^{-3}$. In the simple case obscuration reduces the X-ray flux by a multiplicative factor. If the properties of the obscuring material do not vary significantly on time scales shorter than the high frequency break (e.g., $\Delta t \lesssim 1$ day), then the obscuration should not alter the AGN fractional variability. If that is the case then the $M_{BH}$--$\varsigma^2_X$ relationship should still hold for obscured sources, assuming that their X-ray emission has the same origin as for unobscured sources. Further investigation of the X-ray variability properties of obscured AGN is necessary to confirm this. In addition, it is unclear if the scaling relationship found here applies to AGN radiating at $L / L_{Edd} \lesssim 10^{-3}$. Galactic black holes are observed to undergo a transition in their spectral and variability properties around $L / L_{Edd} \sim 10^{-2}$--$10^{-3}$, likely caused by a change in the accretion flow geometry and state \citep[e.g.,][]{Done2007a}. If supermassive black holes also undergo such state transitions, the variability properties of AGN with $L / L_{Edd} \lesssim 10^{-3}$ may be different than the higher Eddington ratio objects investigated here. Therefore, the scaling relationship presented here should only be used to obtain mass estimates for unobscured AGN with Eddington ratios $10^{-3} \lesssim L / L_{Edd} \lesssim 1$.

Recent Kepler observations suggest that AGN optical PSDs decay as $P(f) \sim f^{-3}$ \citep{Mushotzky2011a}. Because our sup-OU and OU models assume that at high-frequencies $P(f) \sim f^{-2}$, it is of interest to investigate what systematic errors might be introduced by assuming a PSD that is too flat. For a given frequency range, a constant systematic offset is likely introduced to the inferred amplitude of high-frequency variability by assuming a flatter PSD. Because the trends between high-frequency variability and $M_{BH}$ reported here are empirical, any constant offset in $M_{BH}$ caused by the difference between the `true' PSD normalization and that inferred from the RSVP is absorbed into the calibration of the variability-based mass estimates. Therefore, so long as the range of frequencies used to estimate $\varsigma^2$ are similar to the calibration discussed here, there should not be any systematic biased introduced. However, systematic errors may arise when the range of frequencies probed by a lightcurve is different from the lightcurves analyzed in this work.

To assess the impact of mischaracterizing the PSD slope we fit a OU process to the Kepler lightcurve for Zw 229-15 from Q8. This lightcurve is nearly continuously sampled every 30 minutes for 67 days. For the full lightcurve we derive a value of $\varsigma^2_{\rm opt} = 2.96 \pm 0.22 \times 10^{-3}$ fraction$^2$ / sec. We also down-sampled the lightcurve to have a cadence of 1 day; the down-sampled lightcurve samples a range in frequency space that is about two orders of magnitude narrower than the full lightcurve. For the down-sampled lightcurve we derive a value of $\varsigma^2_{\rm opt} = 8.97 \pm 1.89 \times 10^{-3}$ fraction$^2$ / sec. This suggests that for PSDs that decay steeper than $1 / f^2$ the RSVP will overestimate the amplitude of high-frequency in lightcurves that sample narrower frequency ranges. In particular, for a PSD that decays as $P(f) \sim 1 / f^3$ the RSVP will overestimate the high-frequency variability by a factor of $\sim 3$ when the sampled frequency range is reduced by a factor of $\sim 100$, resulting in an underestimate of $M_{BH}$ by a factor of $\sim 1.7$. As it is unlikely that for most applications the frequency ranges probed by AGN lightcurves will vary by more than two orders of magnitude, it is doubtful that systematic errors in $M_{BH}$ larger than a factor of $\sim 2$ will be introduced by incorrectly assuming the OU process model when the true PSD is steeper.

\subsubsection{Comparison with Mass Estimates Derived from Other Scaling Relationships}
\label{s-scaling_relationships}

Mass estimates derived from the various scaling relationships have their advantages and disadvantages. For the host galaxy scaling relationships, they can be used to obtain $M_{BH}$ estimates for both active and inactive galaxies using a single-epoch measurement, so long as the active nucleus can be separated from the host emission. However, this can be difficult to do for bright AGN. Moreover, their primary disadvantage is that the evolution in the scaling relationships, including the amplitude of their scatter, is currently poorly understood. This currently hinders their use as a $M_{BH}$ estimate beyond the local universe.

Mass estimates derived from the broad emission lines have the advantage that they can be estimated from a single-epoch measurement. In addition, they can in principle be employed even at high redshift, as there is good reason to assume that the scaling relationship that they are based on does not exhibit significant evolution. This assumption is based on the fact that the scaling relationship between luminosity and broad line region size is driven by photoionization physics, and should not change with redshift unless certain characteristics of the broad line region change with redshift. The fact that AGN have surprisingly uniform spectra across a broad range of redshift argues that evolution is not a significant source of bias for these mass estimates. One of the disadvantages of these mass estimates is that they are difficult to obtain for AGN with either weak or nonexistent broad emission lines. Another disadvantage is that the scaling relationships involving the \mgii\ and \civ\ lines are not as well understood, and involve extrapolation to regions of the SED parameter space beyond the region occupied by the AGN that they are calibrated with. This is a concern as H$\beta$ redshifts out of the optical at $z \gtrsim 1$, and many investigators turn to \mgii\ and \civ\ at these redshifts. And finally, the line width measurement relies on accurately separating the emission line from the surrounding spectral features, which can be difficult and lead to biases in low $S/N$ spectra \citep[e.g.,][]{Denney2009a,Denney2013a}.

The X-ray RSVP based $M_{BH}$ estimates help to balance out some of the pros and cons of the other scaling relationships. Because the X-ray RSVP scaling relationship has a completely different physical origin than the others, they offer a valuable cross-check against potential systematics. The disadvantages of the $\varsigma_X$-based $M_{BH}$ estimates include the fact that they may not be valid for obscured AGN or those radiating below $L / L_{Edd} \sim 10^{-3}$. The X-ray $\varsigma^2_X$ can be a more expensive measurement than $\sigma_*$ or the broad line mass estimates in terms of observing time. For a $M_{BH} \sim 10^7 M_{\odot}$ AGN the break time scale occurs at $\tau_H \sim 1$ day \citep{Gonzalez-Martin2012a}, and thus the $\varsigma^2_X$ can be measured from a single long-look observation, depending on the orbit of the X-ray observatory. However, at higher masses a monitoring campaign is likely the only way to sample the $1 / f^2$ part of the PSD above the Poisson noise, which is needed in order to measure $\varsigma^2_X$, thus requiring more observations than the single-epoch estimators. However, the X-ray RSVP $M_{BH}$ estimates have several advantages relative to the other scaling relationships. The $\varsigma^2_X$--based mass estimates can be used for AGN regardless of how diluted they are in the optical by their host galaxies. In addition, it is not necessary to calibrate different scaling relationships, as needed for the broad emission lines. And finally, perhaps the strongest advantage is that the X-ray RSVP is a simple and clean quantity to estimate, and does not require separating out the feature of interest through a decomposition, as is necessary to measure the properties of a host galaxy bulge or an emission line width.

It may appear that it is not possible to obtain $M_{BH}$ estimates from $\varsigma_X^2$ for faint AGN or those with only a few epochs in their lightcurve, as the $1 / f^2$ part of the PSD may never appear above the Poisson noise level. In this case the uncertainties on $\varsigma^2_X$ would be so large as to dominate the statistical scatter in the mass estimates. However, even in this case it is still possible to use the X-ray lightcurves for a large sample of AGN to obtain an estimate of the distribution of $M_{BH}$, and consequently the black hole mass function. This is because the variability information in lightcurves for large numbers of AGN can be pooled together to provide meaningful constraints on the distribution of $\varsigma^2_X$, and therefore on $M_{BH}$, even though the values of $\varsigma^2_X$ for any individual AGN may be poorly constrained. Hierarchical modeling provides a statistical framework to estimate the black hole mass function in this situation \citep[e.g.,][]{Kelly2012a}. This is important as it implies that even X-ray monitoring campaigns that only obtain several epochs on time scales $\Delta t \lesssim 1$ month for some area of the sky can still obtain meaningful constraints on the black hole mass function through the ensemble variability, thereby providing an important tool for studying black hole growth. This same argument may apply to the optical lightcurves as well, which, in spite of their larger statistical scatter with respect to $M_{BH}$, may provide meaningful constraints on $M_{BH}$ demographics for large time-domain samples of AGN.

\subsection{Connection with the Accretion Flow}
\label{s-physics}

\begin{figure}
	\includegraphics[scale=0.5,angle=0]{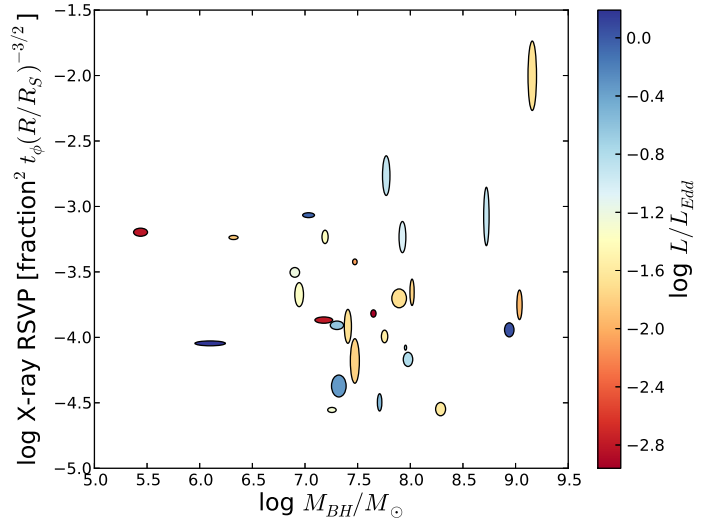}
	\caption{The $M_{BH}$--$\varsigma^2_X$ relationship, expressed in terms of the accretion disk Keplerian frequency, $t_{\phi} = \Omega_{K}^{-1}$. For clarity we have omitted the AGN with only upper limits on $\varsigma^2_X$. When the X-ray RSVP is expressed in terms of fractional variability per disk orbital time scale, there is no evidence for a trend with $M_{BH}$. This suggests that for AGN the fractional variability in the X-rays per unit disk orbital time scale is independent of $M_{BH}$ with an intrinsic scatter of a factor of $\sim 2$.}
	\label{f-mbh_rsvp_torbit}
\end{figure}

In order to lend some insight into what drives the X-ray RSVP correlation with $M_{BH}$, we transform the units of $\varsigma^2_X$ to be with respect to the Keplerian frequency of the accretion disk, $\Omega_K$. The disk orbital time scale is 
\begin{equation}
	t_{\phi} = \Omega_K^{-1} \propto M_{BH} \left(\frac{R}{R_S}\right)^{3/2},
	\label{eq-orbital_time}
\end{equation}
where $R_S$ is the Schwarzschild radius. The dependence of $\varsigma_X^2$ on $M_{BH}$ is shown in Figure \ref{f-mbh_rsvp_torbit}, after transforming $\varsigma^2_X$ to be in units of fractional variance per disk orbital time scale. Because most of the physically relevant time scales in the disk are proportional to $\Omega_k^{-1}$, the dependence of $\varsigma_X^2$ on $M_{BH}$ would be the same if we had used, say, the disk thermal time scale. After transforming $\varsigma_X^2$ to be in units of fractional variance per disk orbital time scale, the correlation with $M_{BH}$ disappears. This suggests that the tight anti-correlation between $\varsigma^2_X$ and $M_{BH}$ is a manifestation of the fact that the fractional variability generated per disk physical time scale on time scales $\Delta t \ll \tau_H$ is independent of $M_{BH}$. The residual scatter in the X-ray RSVP at fixed $M_{BH}$ may reflect variations in the location of the X-ray emitting region in terms of $R_S$, black hole spin, the disk scale height, or the viscosity of the accretion flow. Some of the residual scatter could also be due to deviations in the high-frequency PSD shape from our assumption that $P(f) \propto 1 / f^2$. There is no evidence that the residual scatter in $\varsigma^2_X$ is correlated with the Eddington ratio, implying that the dependence of $\varsigma^2_X$ on $M_{BH}$ is not simply a reflection of the dependence of $\tau_H$ on $M_{BH}$, as $\tau_H$ is correlated with both $M_{BH}$ and $L / L_{Edd}$ \citep[e.g.,][]{McHardy2006a}. Similar conclusions were reached by \citet{Ponti2012a} with regard to the X-ray excess variance.

The optical emission is emitted much further out than the X-ray emission; recent microlensing investigations imply that the optical radiation is generated at $\sim 50$--$100$ gravitational radii, while the X-rays are generated from within several gravitational radii \citep{Morgan2012a}. In addition, the optical emission is thought to be thermal radiation from an optically thick geometrically thin accretion disk, while the X-ray emission comes from an optically thin corona of hot electrons. As seen in Figure \ref{f-rsvp_compare} the amplitude of optical RSVP (corresponding to time scales of $\sim$ days) is uncorrelated with the X-ray RSVP (corresponding to time scales of $\sim $ seconds). This suggests that the optical variability on time scales of $\sim$ days is not primarily driven by reprocessing of X-ray emission, but instead is likely driven by processes in the disk local to where it is emitted.

The optical RSVP is more tightly correlated with luminosity than either $M_{BH}$ or $L / L_{Edd}$. Standard thin accretion disk models assume that the disk is optically thick in the height-direction, implying that the disk should roughly radiate locally as a blackbody. If the accretion disk radiates as a blackbody at a radius $R$, then the disk temperature at $R$ can be derived by equating the dissipation rate per unit face area of the disk to the blackbody flux \citep[e.g.,][]{Frank2002a}:
\begin{equation}
T(R) \propto \left(\frac{M_{BH}\dot{M}}{R^3}\right)^{1/4} \label{eq-t(r)}.
\end{equation} 
Here, $\dot{M}$ is the accretion rate. Under these assumptions, most of the emission at a characteristic wavelength $\lambda$ is coming from a radius
\begin{equation}
R \propto M_{BH}^{1/3} \dot{M}^{1/3} \lambda^{4/3} \label{eq-r}.
\end{equation}
Noting that the disk orbital time scale is $\tau_{\rm orb} \propto R^{3/2} M_{BH}^{-1/2}$ and assuming $L \propto \dot{M}$, we find
\begin{equation}
\tau_{\rm orb} \propto \lambda^2 L^{1/2} \label{eq-torb_lum}.
\end{equation}
Because many other physically relevant time scales are proportional to $\tau_{\rm orb}$, the $L^{1/2}$ scaling holds for a variety of time scales. In particular, for an $\alpha$-disk \citep{Shakura1973a}, the disk thermal time scale is $t_{\rm therm} \propto t_{\rm orb} / \alpha$ where $\alpha$ is the standard viscosity parameter.

Similar to the X-ray RSVP interpretation, if we assume that the fractional optical variability per disk orbital time scale is constant, then Equation (\ref{eq-torb_lum}) implies that $\varsigma^2_{\rm opt} \propto L^{-1/2}$ when the optical RSVP is expressed in units of fractional variability per second. This scaling is in excellent agreement with what we find empirically in Equation (\ref{eq-opt_rsvp_lbol}). Therefore, our empirical trend between the optical RSVP and luminosity is consistent with an interpretation where the disk emits locally as roughly a blackbody, and the fractional variability emitted per disk orbital or thermal time scale is constant. Within this interpretation, the additional scatter in $\varsigma^2_{\rm opt}$ could be caused by variations in the location and size of the region emitting optical radiation at a characteristic wavelength $\lambda$, and by variations in the viscosity $\alpha$. Part of the residual scatter may also be caused by deviations in the high-frequency PSD logarithmic slope from our assumed value of $-2$.

%

\section{SUMMARY}
\label{s-summary}

In this work we investigated correlations involving the amplitude of short time scale AGN optical and X-ray variability and black hole mass, luminosity, and Eddington ratio. In summary:
\begin{itemize}
\item
We presented a new Swift lightcurve for PG 1426+015, the AGN with the highest estimated black hole mass obtained from either reverberation mapping or dynamical modeling, sampling time scales $\approx 2$ days to $\approx 60$ days.
\item
We presented a new Bayesian statistical technique for measuring variability parameters directly from a lightcurve of photon counts for arbitrary sampling. The technique is based on the likelihood function of the measured lightcurve, and thus efficiently uses all of the information in the data. This technique will be useful for measuring variability parameters for faint sources as the X-ray count rates do not need to be binned to obtain Gaussian statistics in the lightcurve; such measurements will enable X-ray variability based black hole mass estimates for AGN over a broader range of luminosities and redshifts.
\item
We analyzed both X-ray and optical archival lightcurves for a sample of 39 AGN, measuring the normalization of the high-frequency part of their PSDs.
\item
We find that the normalization of the high-frequency X-ray PSD is tightly anti-correlated with $M_{BH}$, with no residual trend with Eddington ratio. This anti-correlation is consistent with previous work that has found that the X-ray excess variance is inversely proportional to $M_{BH}$. We also find that this $\varsigma^2_X$--$M_{BH}$ scaling relationship can be used to obtain $M_{BH}$ estimates from the high-frequency X-ray variability with $\approx 0.38$ dex precision relative to the literature values of $M_{BH}$. The $\varsigma^2_X$--$M_{BH}$ trend implies that the high-frequency fractional X-ray variability per disk dynamical time scale is independent of $M_{BH}$ and $L / L_{Edd}$.
\item
We find that the amplitude of optical variability on time scales $\sim$ days is anti-correlated with $M_{BH}, L / L_{Edd},$ and $L$, with the tightest trend being with luminosity. We quantified the trends with $M_{BH}$ and $L$ as $\varsigma^2_{\rm opt} \propto M_{BH}^{-0.52 \pm 0.13}$ and $\varsigma^2_{\rm opt} \propto L^{-0.43 \pm 0.06}$. We find that estimates of $M_{BH}$ based on the optical variability scaling relationship would have a precision of $\sim 0.6$ dex relative to the literature values of $M_{BH}$. While this is not as precise as other scaling relationship, it still may provide reasonable estimates of $M_{BH}$ demographics for large samples of AGN. In addition, the optical variability scaling relationship exhibits an additional trend with luminosity at fixed mass, and therefore it may be possible to obtain $M_{BH}$ estimates with better precision as a function of both $\varsigma^2_{\rm opt}$ and $L$; future work using an unbiased sample of AGN is needed to investigate this. The optical variability trends found here will provide an important foundation for interpreting AGN optical variability in the era of Pan-STARRS and LSST.
\end{itemize}

\section{Acknowledgements}

We would like to thank Aaron Barth, Omer Blaes, Vardha Bennert, Robert Antonucci, and Iossif Papadakis for valuable discussions and comments on our manuscript. We would also like to thank Brad Peterson for generously providing many of the optical lightcurves used in our analysis. BK acknowledges support from the Southern California Center for Galaxy Evolution, a multi-campus research program funded by the University of California Office of Research. BK, TT, and AP acknowledge support from the Packard foundation through a Packard Research Fellowship to TT. JHW acknowledges support form the National Research Foundation of Korea (NRF) grant funded by the Korea government (No.2012-006087). This research has made use of the XRT Data Analysis Software (XRTDAS) developed under the responsibility of the ASI Science Data Center (ASDC), Italy.

\end{document}